\documentclass[a4paper,
english,
11pt,
DIV11]{scrartcl}

\usepackage{float}
\usepackage{textcomp}
\usepackage{latexsym} 
\usepackage{amsfonts} 
\usepackage{amssymb}
\usepackage[cmex10]{amsmath}
\usepackage{amsthm}
\usepackage{stmaryrd}
\usepackage[T1]{fontenc}
\usepackage[utf8]{inputenc}
\usepackage{type1cm}
\usepackage{hyperref}
\usepackage{enumerate}
\usepackage{url}
\usepackage{pgf}
\usepackage{tikz}
\usepackage{slashed} 
\usepackage{xargs}
\usepackage{xifthen}
\usepackage{MnSymbol}
\usepackage{bbding}
\usepackage{comment}
\usepackage{pop}
\usepackage{editing}
\usepackage{multirow}
\usepackage[numbers,sort&compress]{natbib}

\theoremstyle{plain}
\newtheorem{theorem}{Theorem}
\newtheorem{lemma}[theorem]{Lemma}

\newtheorem{corollary}[theorem]{Corollary}
\newtheorem{definition}[theorem]{Definition}

\newtheorem{claim}[theorem]{Claim}

\deffootnote{1em}{1em}{\textsuperscript{\thefootnotemark\ }}

\newenvironment{IEEEproof}[1][]{\begin{proof}\bgroup}{\egroup\end{proof}}

\title{A New Order-theoretic Characterisation of the
Polytime Computable Functions%
\thanks{This work is partially supported by FWF (Austrian Science Fund) project 
I-608-N18 and by a grant of the University of Innsbruck.}}

\author{Martin Avanzini${}^{1}$, Naohi Eguchi${}^{2}$ and Georg Moser${}^{1}$\\
${}^{1}$ Institute of Computer Science,
University of Innsbruck,
Austria \\[-1mm]
{\small\texttt{\{martin.avanzini,georg.moser\}@uibk.ac.at}}\\
${}^{2}$ Mathematical Institute
Tohoku University,
Japan \\[-1mm]
{\small\texttt{\{eguchi@math.tohoku.ac.jp}}
}

\begin{document}
\maketitle

\begin{abstract}
We propose a new order, the \emph{small polynomial path order} 
(\emph{\POPSTARS} for short). The order \POPSTARS\ 
provides a characterisation of the class of polynomial time computable function
via term rewrite systems.
Any polynomial time computable function
gives rise to a rewrite system that is compatible with \POPSTARS. 
On the other hand any function defined by a rewrite system 
compatible with \POPSTARS\ is polynomial time computable. 

Technically \POPSTARS\ is a tamed recursive path order with product status. 
Its distinctive feature is the precise control provided. 
For any rewrite system that is compatible with \POPSTARS\ that makes use of 
recursion up to depth $d$, the (innermost) runtime complexity 
is bounded from above by a polynomial of degree $d$.
\end{abstract}

\section{Introduction}\label{s:intro}

In this paper we are concerned with the \emph{complexity analysis} of term rewrite
systems (TRSs) and the ramifications of such an analysis in 
\emph{implicit computational complexity}. 

Term rewriting is a conceptually simple, but powerful abstract model of computation. 
The foundation of rewriting is equational logic and 
\emph{term rewrite systems} (\emph{TRSs} for short) 
are conceivable as sets of directed equations.
The implicit orientation of equations in TRSs naturally 
gives rise to computations, where a term is rewritten by successively replacing
subterms by equal terms until no further reduction is possible. 
Such a sequence of rewrite steps is also called a \emph{derivation}.

A natural way to measure the complexity of a TRS $\RS$ is to measure the
length of computations in $\RS$. More precisely the \emph{runtime complexity} 
of a TRS relates the maximal lengths of derivations to the size of the initial
term. Furthermore the shape of the initial term is suitable restricted. The
latter restrictions aims at capturing the complexity of the 
\emph{functions computed} by the analysed TRS. 
Indeed, the runtime complexity of a TRS $\RS$ forms an 
\emph{invariant cost model}. Suppose the runtime complexity of $\RS$
is polynomially bounded and the function computed by $\RS$ is implemented
on a Turing machine. Then the runtime of this Turing machine is polynomially
bounded~\cite{AM10}.

We propose a new order, the \emph{small polynomial path order} 
(\emph{\POPSTARS} for short). The order \POPSTARS\ 
provides a characterisation of the class of 
\emph{polynomial time computable function} 
(\emph{polytime computable functions} for short) via term rewrite systems.
Any polytime computable function
gives rise to a rewrite system that is compatible with \POPSTARS. 
On the other hand any function defined by a rewrite system 
compatible with \POPSTARS\ is polytime computable. 
The proposed order embodies the principle of \emph{predicative recursion} as
proposed by Bellantoni and Cook~\cite{BC92}. 
Our result bridges the subject of (automated) complexity analysis of
rewrite systems and the field of implicit computational complexity 
(\emph{ICC} for short).

Our results entail a new syntactic criteria to automatically
establish polynomial runtime complexity of a given TRS.
This criteria extends the state of the art in runtime complexity 
analysis as it is more precise or more efficient than related techniques. 
Note that the analysis is automatic as for any given TRS, compatibility with
\POPSTARS\ can be efficiently checked by a machine. Should this check succeeds
we get an asymptotic bound on the runtime complexity directly from the parameters
of the order. 
It should perhaps be emphasised that compatibility
of a TRS with \POPSTARS\ implies termination and thus our complexity
analysis technique do not presuppose termination, instead
we use (variations) of termination techniques to induce upper bounds on
the complexity.

Our syntactic account of predicative
recursion delineates a class of rewrite systems: a rewrite system $\RS$
is called \emph{predicative recursive of degree $d$} if $\RS$ is
compatible with \POPSTARS\ and the depth of recursion of all 
function symbols in $\RS$ is bounded by $d$ (see Section~\ref{s:spopstar} for the formal
definition). Any predicative recursive rewrite system of degree $d$ 
admits runtime complexity in $O(n^d)$.

\subsection{Related works}

Polynomial runtime complexity analysis is an active research area
in rewriting. Interest in this field greatly increased recently. This is
partly due to the incorporation of a dedicated category for complexity 
into the annual termination competition (TERMCOMP).%
\footnote{\url{http://termcomp.uibk.ac.at/}.} 
We mention very recent work on matrix interpretations that is readily applicable
to runtime complexity analysis by Middeldorp et al.~\cite{MMNWZ11} and recent 
work on the incorporation of the dependency pair method in complexity 
analysis~\cite{HM08,NEG11,HM11}. See~\cite{M09}
for an overview on work on complexity analysis in rewriting.
The most powerful techniques for runtime complexity analysis currently available,
basically employ semantic considerations on the rewrite systems, which are notoriously
inefficient.

There are several accounts of predicative analysis of recursion in the
(ICC) literature. We mention only those related works which are directly
comparable to our work. See~\cite{BMR09} for an overview on ICC. 
Notable the clearest connection of our work is to Marion's \emph{light multiset path order}
(\emph{LMPO} for short)~\cite{Marion03}. This path order forms a strict extension of the here 
proposed order \POPSTARS, but lacks the precision of the latter. 
In Bonfante et.~al.~\cite{BCMT01} restricted classes of polynomial
interpretations are studied that can be employed to obtain similar 
precise polynomial upper bounds on the runtime complexity of TRSs as with
\POPSTARS. Neither of these results is applicable to relate the depth
of recursion to the runtime complexity, in the sense mentioned above.
We have also drawn motivation from~\cite{Marion09} which provides a related
fine-grained capturing of the polytime computable functions, but which lacks
applicability in the context of runtime complexity analysis.

Above we emphasised that our analysis is automatic and there are other
recent approaches for the automated analysis of resource usage in programs.
Notable Hoffmann et al.~\cite{HAH11} provide an automatic multivariate amortised 
resource analysis which extends earlier results on an automatic cost analysis using typing.
Albert et al.~\cite{AAGGPRRZ:2009} present an automated complexity tool
for Java Bytecode programs and
Gulwani et al.~\cite{GMC09} provide an  automated complexity tool for C programs.

\subsection{Contributions}
\subsubsection{Precise Runtime Complexity Analysis}

The proposed order \POPSTARS\ is a restriction of the
\emph{polynomial path order} (\emph{\POPSTAR} for short) introduced
by the second and third author~\cite{AM08}. Crucially \POPSTARS\ is a 
tamed recursive path order with product status~\cite{BN98}. 
Its distinctive feature is the precise control provided for
runtime complexity analysis: for any predicative recursive
TRS of degree $d$ its runtime complexity lies in $O(n^d)$
(cf.~Theorem~\ref{t:spopstar}).
Furthermore this bound is tight, that is, we provide a family of TRSs, 
delineated by \POPSTARS, whose runtime complexity is bounded from below by $\Omega(n^d)$.

\subsubsection{Fine-Grained Capture of Polytime Functions}

As already mentioned the runtime complexity of a TRS forms an invariant cost
model. Hence \POPSTARS\ is \emph{sound} for the class of polytime computable
functions: any function $f$ computable by a TRS $\RS$, such that $\RS$
is compatible with \POPSTARS\ is polytime computable. On the other hand
\POPSTARS\ is \emph{complete}: for any polytime computable function $f$ 
there exists a TRS $\RS$ computing $f$ such that $\RS$ is compatible with \POPSTARS.
However, for runtime complexity, we can obtained a more fine-grained
classification. We establish that those TRS that are definable
with $d$ nestings of predicative recursion are 
predicative recursive of degree $d$ (cf.~Theorem~\ref{thm:strong_completeness}). 
Conclusively the runtime complexity of these systems lies in $O(n^d)$.
Thereby we obtain a fine-grained characterisation
of the polytime computable functions, which may be of interest in
implicit computational complexity theory.



\subsubsection{Parameter Substitution}

We extend upon \POPSTARS\ by proposing 
a generalisation of \POPSTARS, admitting the same properties as above, 
that allows to handle more general recursion schemes that make
use of parameter substitution (cf.~Theorem~\ref{t:spopstarps}).
As a corollary to this and the fact that the runtime complexity
of a TRS forms an invariant cost model we conclude a non-trivial
closure property of the class $\Bwsc$: this
class is closed under predicative recursion with parameter substitution. 

\subsubsection{Automated Complexity Analysis }

We have implemented the order \POPSTARS\ in the 
\emph{Tyrolean Complexity Tool} \TCT, version 1.9, 
an open source complexity analyser.%
\footnote{Available at \url{http://cl-informatik.uibk.ac.at/software/tct}.}
The experimental evidence obtained indicates the viability of the method.

\section{Motivation}

We present the main ideas of the proposed \emph{small polynomial path order} and provide
an informal account of the technical results obtained in the remainder of the paper.

The order \POPSTARS\ essentially embodies the predicative analysis of recursion
set forth by Bellantoni and Cook. In~\cite{BC92} a recursion-theoretic
characterisation $\mathcal{B}$ of the class of polytime computable functions is 
proposed. This analysis is connected to the important principle of \emph{tiering}
introduced by Simmons~\cite{Simmons:1988} and Leivant~\cite{L91}. The essential
idea is that the arguments of a function are separated into \emph{normal} 
and \emph{safe} arguments (or correspondingly into arguments of different tiers). 

It is a natural idea to seek out term-rewriting characterisations of the
polytime computable functions~\cite{BW96}. Indeed, Beckmann and Weiermann 
successfully apply their characterisation to yield a 
non-trivial closure property of the class $\mathcal{B}$. 
Given a TRS $\RS$ representing a polytime computable function, we
seek syntactic criteria on $\RS$ to 
verify that the runtime complexity of $\RS$ lies 
in $O(n^d)$ for $d \in \N$.

Let us make this idea precise. We present a subclass $\Bwsc$ of
$\mathcal{B}$ that only induces TRSs compatible with \POPSTARS. 
We formulate the class $\Bwsc$ over the set $\{0,1\}^\ast$ of binary
words, where we write $\epsilon$ to denote the empty sequence and
$S_i(;x)$ to denote the word $xi$.
We are assuming that the arguments of every function are
partitioned in to normal and safe ones. 
Notationally we write $f(\pseq[k][l]{t})$ where argument are separated 
by a semicolon. Normal arguments are always drawn to the left, 
and safe arguments to the right of the semicolon.
The class $\Bwsc$ is the smallest class containing
certain initial functions and closed under 
\emph{weak safe composition} and  
\emph{safe recursion on notation}, which are presented in
Fig.~\ref{fig:Bwsc}.
Due to a variation of a result by Handley and Wainer, 
we have that $\Bwsc$ captures the polytime functions~\cite{HW99}.
A term-rewriting characterisation of the polytime functions is 
obtained by orienting the equations in Fig.~\ref{fig:Bwsc} from left to right.

\begin{figure*}
\centering
\begin{tabular}{l@{\hspace{3mm}}lrl}
  \textbf{Initial Functions}
  & $P(;\epsilon) =  \epsilon$ \\
  & $P(; S_i(;x)) =  x$ &  $(i=0,1)$ \\
  & $I^{k,l}_j(\vec{x};\vec{y}) =  x_j$ & ($j \in \{1,\dots,k \}$) \\
  & $I^{k,l}_j(\vec{x};\vec{y}) = y_{j-k}$ & ($j \in \{k+1, \dots, l+k\}$) \\
  & $C(;\epsilon, y, z_0, z_1) = y$ \\
  & $C(;S_i(;x), y, z_0, z_1) = z_i$ & ($i = 0, 1$) \\
  & $O(\vec{x};\vec{y}) = \epsilon$ 
  \\[3mm]
  \textbf{Weak Composition} ($\m{WSC}$)
  & \multicolumn{3}{@{}l}{$f(\vec x; \vec y) = h(x_{i_1},\dots,x_{i_k};\vec{g}(\vec{x};\vec{y}))$} 
  \\[3mm]
  \textbf{Safe Recursion} ($\m{SRN}$) 
  & \multicolumn{3}{@{}l}{$f(\epsilon, \vec x; \vec y) =  g(\vec x; \vec y)$}
  \\
  & \multicolumn{3}{@{}l}{$f(S_i (; z), \vec x; \vec y) = h_i (z, \vec x; \vec y, f(z, \vec x; \vec y))$ ($i = 0, 1$)}
\end{tabular}
\caption{Defining initial functions and operations for $\Bwsc$}
\label{fig:Bwsc}
\end{figure*}

Suppose the definition of $\RS$ is based on the equations in $\Bwsc$. 
It seems likely to deduce a precise bound on the runtime complexity of $\RS$
by measuring the number of nested applications of safe recursion. 
We establish that such a TRS is predicative recursive of degree $d$, where
$d$ is the maximal nesting of schema ($\m{SRN}$) 
(cf.~Theorem~\ref{thm:strong_completeness}).
This result is based on a suitable definition of \POPSTARS: the parameters
and order constraints present reflect the operators in
the class $\Bwsc$. 

In order to employ the separation of normal and safe arguments, 
we fix for each defined symbol
a partitioning of argument positions into \emph{normal} and \emph{safe} positions. 
For constructors we fix that all argument positions are safe. 
Moreover \POPSTARS\ restricts recursion to normal argument.
Dually only safe argument positions allow the substitution of recursive calls.
Via the order constraints we can also guarantee that only normal arguments are
substituted for normal argument positions. 
Hence \POPSTARS\ enforces a weak composition
schema for function composition, as well as a safe recursion schema for
recursion. For the latter the comparison of arguments via a product
status, rather than via a multiset status is also essential.
The formal definition of \POPSTARS\ is given in Section~\ref{s:spopstar}.

We remark on the connection between \POPSTARS\ and other path orders.
It is clear that an order-theoretic characterisation 
of predicative recursion is obtained as a restriction of the
recursive path order. Predicative recursion stems from a careful
analysis of primitive recursion and the recursive path order (with multiset status)
characterises the class of primitive recursive functions~\cite{H92}. 
The light multiset path order proposed by Marion is 
based on the separation of normal and safe arguments and adapts function composition 
suitable to retain the separation of normal and arguments. However, the composition
schema goes beyond weak composition. It is shown in~\cite{Marion03} that
LMPO captures the polytime computable functions, but this result relies 
on a clever use of memoisation techniques. In particular the aforementioned restrictions
are not enough to forbid the treatment of TRSs that 
induce non-feasible runtime complexity.
To recover the situation, \POPSTAR~additionally requires the
absence of multiple recursive calls in a rule. 
Then it can be shown that in fact compatible 
TRSs admits feasible runtime complexity~\cite{AM08}. However, for
rewrite systems of predicative recursion of degree $d$, \POPSTAR\ 
overestimates the runtime complexity.

\begin{figure}
\centering
\begin{tikzpicture}
    \newcommand{\n}{2};
    \newcommand{\dx}{1.8};
    \newcommand{\dy}{0.9};
    \newcommand{\drawterms}[2]{
      \pgfmathparse{#1}
      \node (s_#1) at (\pgfmathresult*\dx,0) {$s_{#2}$};
      \node (is_#1) at (\pgfmathresult*\dx,-\dy) {$\ints(s_{#2})$};
      \draw[->] (s_#1) to node[right] {} (is_#1); 
    }
    \newcommand{\drawrel}[1]{
      \pgfmathparse{#1}
      \node (rew_#1) at (\pgfmathresult*\dx+\dx/2,0) {$\irew$};
      \node (ord_#1) at (\pgfmathresult*\dx+\dx/2,-\dy) {$\gspopv$};
    }

    \foreach \x in {1,...,\n} \drawterms{\x}{\x};

    \pgfmathparse{(\n+1)*\dx};
    \node at (\pgfmathresult,0) {$\dots$};
    \node at (\pgfmathresult,-\dy) {$\dots$};
    \drawterms{\n+2}{\ell};

    \foreach \x in {1,...,\n} \drawrel{\x};
    \drawrel{\n+1};
  \end{tikzpicture}
  \caption{Embedding of the innermost rewrite relation into $\gspopv$.}
  \label{fig:embed}
\end{figure}

To show that \POPSTARS\ is correct, 
we make use of a variety of ingredients. 
First in Section~\ref{s:spopstar}
we propose a family of TRSs that establish the tightness of the obtained bound. 
Then in Section~\ref{s:pint}, we define \emph{predicative interpretations} $\ints$
that flatten terms to \emph{sequences of terms}, essentially separating 
safe from normal arguments.
This allows us to analyse a term independent from its safe arguments.
In Section~\ref{s:spop} we introduce an order $\gspopv$ on sequences of terms, 
that is simpler compared to $\gspop$ and does not rely on the separation
of argument positions.  
In Section~\ref{s:embed} we show that predicative interpretations $\ints$ 
embeds innermost rewrite steps into $\gspopv$ as depicted in Fig.~\ref{fig:embed}.
As the length of $\gspopv$ descending sequences
starting from basic terms can be bound appropriately (cf. Theorem~\ref{t:spop}),
we obtain correctness.
Finally, to show that \POPSTARS\ is complete, it suffices to show that 
any $\Bwsc$ only induces TRSs 
compatible with \POPSTARS (cf.~Section~\ref{s:completeness}).

\section{Preliminaries}

We assume at least nodding acquaintance with the basics
of rewriting cf.~\cite{BN98}.
In this section we fix the bare essential of notions and notation, we use
in the remainder of the paper.

Throughout the paper, we fix a countably infinite set of \emph{variables} $\VS$
and a finite \emph{signature} $\FS$ consisting of \emph{function symbols} $f,g,h, \dots$.
The signature associates with each function symbol $f$ a natural number, its \emph{arity}.
The set of \emph{terms} is denoted as $\TERMS$ (or $\TA$ for short).
We write $s \superterm t$ to indicate that  $s$ is a \emph{subterm} of $t$ and
$s \supertermstrict t$, if $s$ is a \emph{proper subterm}.
The \emph{size} of $t$ refers to the number of function symbols 
and variables it contains. 
The \emph{depth} $\depth(t)$ of $t$ 
is $0$ if $t$ is a variable or a constant, 
for $t = f(\seq{t})$ the depth of $t$ is $d + 1$ where
$d$ is the maximal depth of an argument $\seq{t}$.

A precedence $\qp$ is a preorder on the set of function symbols $\FS$.
As usual $\qp$ induces an equivalence $\ep$ and a strict proper order $\sp$.
If $f \sp g$ we say that $g$ is \emph{(strictly) below} $f$ in the precedence.
This will indicate that the definition of $f$ depends on $g$.
In this sense $\qp$ can be seen as a form of static call graph.
The \emph{rank} $\rk(f) = 1 + \max \{\rk(g) \mid f \sp g\}$ 
measures the length of $\sp$ chains starting from the function symbol $f$. 
(We employ the convention that the maximum of an empty set equals $0$.)
The equivalence $\ep$ is lifted from function symbols to 
terms by additionally disregarding the order on arguments.
Formally $s$ and $t$ are \emph{equivalent}, in notation $s \eqi t$,
if $s = t$, or $s = f(\seq{s})$ and $t = g(\seq{t})$ where
$f \ep g$ and $s_i \eqi t_{\pi(i)}$ for all arguments
and some permutation $\pi$.

A \emph{rewrite rule} is a pair $(l,r)$ of terms, denoted 
as $l \to r$, where the \emph{left-hand side} $l$ is not a variable
and the \emph{right-hand side} $r$ mentions only variables from $l$.
A \emph{term rewrite system} (\emph{TRS} for short) $\RS$ 
(over the signature $\FS$) is a \emph{finite} set of rewrite.
The root symbols of left-hand sides are called \emph{defined symbols}, 
the remaining function symbols are called \emph{constructors}.
A term that only contains constructors is also called \emph{value}, 
and the set of all values is denoted by $\Val$.

Let $\RS$ denote a TRS.
The TRS $\RS$ induces the \emph{rewrite relation} $\rew$ on terms as follows.
Informally, a term $s$ \emph{rewrites} to $t$ if 
the left-hand side $l$ of a rewrite rule $l \to r$ from $\RS$
matches some subterm of $s$, and the term $t$ is obtained from $s$ 
by replacing the matched subterm with the corresponding instance of the right-hand side $r$.
Formally, $s \rew t$ if there exists
a \emph{context} $C$, \emph{substitution} $\sigma$ and rule ${l \to r} \in \RS$
such that $s = C[l\sigma]$ and $t=C[r\sigma]$.
Here a context $C$ is a term with exactly one occurrence 
of a \emph{hole} $\hole$, and 
$C[t]$ denotes the term obtained by replacing the hole $\hole$ 
in $C$ by $t$.
A substitution $\sigma$ is a function that maps 
variables to terms, and $t\sigma$ denotes
the homomorphic extension of this function to terms.

A term $t$ is called a \emph{normal form} (with respect to $\RS$)
if it is irreducible, ie., if there exists no term $s$ 
with $t \rew s$. Consider the rewrite step 
$s \rew t$ where $s = C[l\sigma]$ as above.
If all arguments of $l\sigma$ are normal forms, 
then this step is an \emph{innermost rewrite step} and denoted 
by $s \irew t$. 
The relation $\irew$ is called the 
\emph{innermost rewrite relation} of $\RS$.
It requires arguments to be evaluated first, 
and can be seen as the adoption of a call-by-value semantics.
In the sequel we will only be concerned with innermost rewriting.

The TRS $\RS$ is a \emph{constructor TRS} if arguments of left-hand
sides only contain constructors. 
It is \emph{completely defined} if values coincide with normal forms, that is, 
defined symbols do not occur in normal forms.
The TRS $\RS$ is called \emph{terminating} if $\rew$ is well-founded, 
and $\RS$ is \emph{confluent} if all peaks
$t_1 \mathrel{{}_\RS^{~*}{\leftarrow}} s \rightarrow^{*}_\RS t_2$ can be joined:
$t_1 \rightarrow^{*}_\RS u \mathrel{{}_\RS^{~*}{\leftarrow}} t_2$ for some term $u$.
If a TRS is confluent and terminating, then the result $t_\ell$
of a \emph{computation} of $s$, $s \rew t_0 \to \dots \rew t_\ell$ where
$t_\ell$ is in normal form, is well-defined
and unique.
Hence this subclass of TRSs forms a model of deterministic computation.
Note that the model is independent on the evaluation strategy. 

In this paper we are interested in the \emph{runtime complexity} of such computations, 
following \cite{HM08} we measure runtime as the number of rewrite
steps in relation to input sizes, and disregard starting terms that do not
correspond to function calls:
The set of \emph{basic terms} $\Tb$ constitutes of all 
terms $f(\vec{v})$ where $f$ is defined and the arguments $\vec{v}$ are values.
The \emph{(innermost) runtime complexity} of a terminating TRS $\RS$ is defined 
as
$
\rcsym^{\text{\scriptsize{(\textsf{i})}}}_{\RS}(n) 
\defsym \max \{\dheight(t,\to) \mid \text{$t$ is a basic of size up to $n$}\}$.
Here $\to$ denotes $\rew$ or $\irew$ respectively, and the \emph{derivation height}
$\dheight(t,\to)$ is the maximal length of a derivation starting in $t$.
As we focus in paper on innermost rewriting, we often drop the qualification
\emph{innermost}, when referring to the runtime complexity of a TRS. No confusion
will arise from this.


\section{The Small Polynomial Path Order}\label{s:spopstar}

We arrive at the definition of \POPSTARS.
To precisely assess the complexity of a TRS, \POPSTARS\ 
allows recursive definitions only on a subset of defined symbols, 
the so called \emph{recursive symbols}. Symbols that are not 
recursive are called \emph{compositional}.

Let $\RS$ be a TRS and fix a precedence $\qp$ on the symbols of $\RS$.
To assert our understanding that $\qp$ reflects a call graph 
indifferent on equivalent symbols, we require that $\qp$ is \emph{admissible}:
(i) constructors do not depend on defined symbols: $f \sp g$ implies that $f$ is not a constructor, and
(ii) the equivalence $\ep$ adheres the separation of constructors, recursive and compositional symbols:
if $f \ep g$ then both $f$ and $g$ are either constructors, recursive or compositional symbols.
The \emph{depth of recursion} $\recdepth(f)$ 
is defined in correspondence to the rank $\rk(f)$, but only takes recursive
symbols into account:
\begin{equation*}
  \recdepth(f) \defsym
  \begin{cases}
    1+\max\{\recdepth(g) \mid f \sp g\} & \text{ if $f$ is recursive, and}\\
    \max\{\recdepth(g) \mid f \sp g\} & \text{ otherwise \tpkt}
  \end{cases}
\end{equation*}

\POPSTARS\ does not differentiate between equivalent terms in principle, 
however the equivalence need to respect 
the separation of normal and safe argument positions.
We formalise this in the equivalence relation $\eqis$, where
$s \eqis t$ holds if
 $s = t$ or
 $s = f(\pseq[k][l]{s})$, $t = g(\pseq[k][l]{t})$ where
$f$ is equivalent to $g$ and $s_i \eqis t_{\pi(i)}$ for all argument positions $i = 1,\dots,k+l$.
Here $\pi$ denotes a permutation on argument positions so that position $\pi(i)$ is
normal if and only if the position $i$ is normal.
Let $s = f(\pseq[k][l]{s})$ we define the relation $\nsubtermstrict$ 
so that $s \nsubtermstrict t$ holds if
(i) $s_i \eqis t$ or $s_i \nsubtermstrict t$ for some argument $s_i$ of $s$, and 
(ii) if $f$ is defined then the argument position $i$ is normal ($i \in \{1,\dots,k\}$).
In any case $s \nsubtermstrict t$ implies that $t$ is equivalent to a subterm of $s$.

The following definition introduces small polynomial path orders $\gspop$.
\begin{definition}\label{d:gspop}
  Let $s$ and $t$ be terms such that $s = f(\pseq[k][l]{s})$.
  Then $s \gspop t$ if one of the following alternatives holds.
  \begin{enumerate}
    \item\label{d:gspop:st} $s_i \geqspop t$ for some argument $s_i$ of $s$.
    \item\label{d:gspop:ia} $f$ is a defined symbol, $t = g(\pseq[m][n]{t})$ 
      such that $g$ is below $f$ in the precedence and
      the following conditions hold:
      \begin{enumerate}
      \item\label{d:gspop:ia:1} $s \nsubtermstrict t_j$ for all normal arguments $t_j$ of $t$;
      \item\label{d:gspop:ia:2} $s \gspop t_j$ for all safe arguments $t_j$ of $t$;
      \item\label{d:gspop:ia:3} at most one argument $t_j$ of $t$ contains 
        defined symbols not below $f$ in the precedence.
      \end{enumerate}
    \item\label{d:gspop:ts} $f$ is recursive and $t = g(\pseq[k][l]{t})$
      such that $g$ is equivalent to $f$ in the precedence and the following conditions hold:
      \begin{enumerate}
      \item $\tuple{s_1,\dots,s_k} \gspop \tuple{t_{\pi(1)},\dots,t_{\pi(k)}}$ 
        for some permutation $\pi$ on normal argument positions;
      \item $\tuple{s_{k+1},\dots,s_{k+l}} \gspop \tuple{t_{\tau(k+1)},\dots,t_{\tau(k+l)}}$ 
        for some permutation $\tau$ on normal argument positions.
      \end{enumerate}
    \end{enumerate}
  Here $s \geqspop t$ denotes that either $s$ and $t$ are equivalent or 
  $s \gspop t$ holds. 
  In the last clause we use $\gspop$ also for the product 
  extension of $\gspop$, where
  $\tup[n]{s} \geqspop \tup[n]{t}$ means 
  $s_i \geqspop t_{i}$ for all $i = 1,\dots,n$, 
  and $\tup[n]{s} \gspop \tup[n]{t}$ indicates that additionally 
  $s_{i_0} \gspop t_{i_0}$ holds for at least one $i_0 \in \{1,\dots,n\}$.
\end{definition}

We say that a TRS $\RS$ is \emph{compatible} with $\gspop$ if all rules are \emph{oriented}
from left to right: $l \gspop r$ for all rules ${l \to r} \in \RS$.
We use the notation \caseref{\gspop}{i} to refer to the \nth{$i$} case in 
Definition~\ref{d:gspop} (a similar notation is employed for the subsequently defined orders).

\paragraph*{Some Comments on the Definition} 
Consider a compatible TRS $\RS$.
By compatibility the left-hand side $l$ is compared to the right-hand side of $r$
and, recursively the arguments of $r$, for each rule $l \to r$ of $\RS$.
The case \cref{gspop}{st} is standard in recursive orders and allows the treatment
of functions defined by projection.
Consider the more involved cases where the orientation is due to 
$\cref{gspop}{ia}$ or $\cref{gspop}{ts}$.
We use case \cref{gspop}{ia} to capture function composition, where 
$f$ is defined in terms of a function $g$ below $f$ in the precedence.
The order constraints on normal arguments
enforce that safe arguments of $f$ cannot pass to normal arguments of $g$, 
and moreover the use of $\nsubtermstrict$ disallows composition in 
normal positions of $g$.
In contrast, safe arguments of the right-hand side can be compared using the full power of $\gspop$. 
The only additional restriction imposed states that at most one recursive
call can occur below the function symbol $g$.
Finally the case \cref{gspop}{ts} 
captures recursion and is employed when $l$ is compared to the recursive call.
Here we require that the product of arguments decrease, where
we are careful not to mix normal and safe arguments.
In addition we require that normal arguments, ie. the recursion parameters, decrease strictly
between $l$ and the recursive call in $r$.

We say a constructor TRS $\RS$ is \emph{predicative recursive of degree $d$} if
$\RS$ is compatible with an instance $\gspop$ and the maximal depth of 
recursion of a function symbol in $\RS$ is $d$.

\begin{theorem}\label{t:spopstar}
  Let $\RS$ be predicative recursive of degree $d$.
  Then the innermost derivation height of any basic term 
  $f(\svec{u}{v})$ is bounded by a polynomial of degree $d$ in the 
  sum of the depths of normal arguments $\vec{u}$.
\end{theorem}

As corollary to Theorem~\ref{t:spopstar} we obtain 
that $\gspop$ induces polynomial innermost runtime complexity 
on constructor TRSs.

\begin{corollary}\label{c:spopstar}
If $\RS$ is predicative recursive of degree $d$, then 
the innermost runtime complexity of $\RS$ lies in $O(n^d)$.
\end{corollary}

Consider the constructor TRS $\RSsquare$, whose
rules are given in Fig.~\ref{fig:square}.
The TRS $\RSsquare$ defines squaring of natural numbers
build from the constructors $\mZ$ and $\mS$.
Consider the precedence so that
${\mSquare} \sp {\mTimes} \sp {\mPlus} \sp {\mS} \ep {\mZ}$.
Then it can be verified that the TRS $\RSsquare$ is compatible with $\gspop$.
For instance $\mTimes(\sn{\mS(\sn{}{x}), y}{}) \gspop \mPlus(\sn{y}{\mTimes(\sn{x,y}{})})$
follows by \cref{gspop}{ia} since $y$ appears as normal argument in the left-hand side
and $\mTimes(\sn{\mS(\sn{}{x}), y}{}) \gspop \mTimes(\sn{x,y}{})$
follows by one application of \cref{gspop}{ts}.
Note that the orientation only requires addition ($\mPlus$) and multiplication ($\mTimes$) 
to be recursive symbols, but not the square function ($\mSquare$).
Hence the precedence gives a recursion depth of $2$ for multiplication and squaring, 
and a recursion depth of $1$ to addition.
According to Theorem~\ref{t:spopstar} addition gives rise to linear, 
and multiplication as well as squaring gives rise to 
quadratic runtime complexity.
Overall, the runtime complexity is quadratic.
\begin{figure}
  \centering
  \begin{alignat*}{4}
    \mPlus(\sn{\mZ}{y}) & \to y &
    \mTimes(\sn{\mZ, y}{}) & \to \mZ \\
    \mPlus(\sn{\mS(\sn{}{x})}{y}) & \to \mS(\mPlus(\sn{x}{y})) \quad &
    \mTimes(\sn{\mS(\sn{}{x}), y}{}) & \to \mPlus(\sn{y}{\mTimes(\sn{x,y}{})}) \\
    \mSquare(\sn{x}{}) & \to \mTimes(\sn{x}{x})
  \end{alignat*}  
  \caption{Rewrite system $\RSsquare$}
  \label{fig:square}
\end{figure}

We emphasise that Corollary~\ref{c:spopstar} is tight in the sense
that for any $d \in \N$ there exists a predicative recursive
TRS of degree $d$ so that the runtime complexity 
is bounded from below by $\Omega(n^d)$.

To see this, define a family of TRSs $\RS_i$ ($i\in\N$) inductively as follows:
$\RS_0 \defsym \{ \m{f}_0(\sn{x}{}) \to \m{a} \}$ and
$\RS_{i+1}$ extends $\RS_i$ by the rules presented in Fig.~\ref{fig:ext}.
\begin{figure}
  \centering
  \begin{alignat*}{2}
      \m{f}_{i+1}(\sn{x}{}) & \to \m{g}_{i+1}(\sn{x,x}{}) \\
      \m{g}_{i+1}(\sn{\m{s}(\sn{}{x}),y}) 
      & \to \m{b}(\sn{}{\m{f}_{i}(\sn{y}{}),\m{g}_{i+1}(\sn{x,y})}) \tpkt
  \end{alignat*}  
  \caption{Extension rules of TRS $\RS_{i+1}$, where
    $\m{f}_{i+1}$ and $\m{g}_{i+1}$ are fresh}
  \label{fig:ext}
\end{figure}
By construction $\RS_d$ is compatible with $\gspop$ as induced by the 
precedence
$
\m{f}_{d} \sp \m{g}_{d} \sp 
\m{f}_{d-1} \sp \m{g}_{d-1} \sp 
\dots \sp
\m{f}_0 \sp \m{a} \ep \m{b}$, where
only the defined symbols $\m{g}_{i}$ ($i = 1,\dots,d$)
are recursive. Obviously the maximal depth of recursion of $\RS_d$ is $d$

We show that the runtime complexity of $\RS_d$ is in $\Omega(n^d)$:
For $d = 0$ this is immediate. 
For $d > 1$, note that $\m{g}_{d}$ performs recursion
on its first argument, at each step calling $\m{f}_{d -1}$.
Conclusively $\m{f}_{d}(\m{s}^n(\m{a}))$ calls $n$ times the
function $\m{f}_{d -1}$. 
Inductive reasoning yields that 
$\m{f}_{d}(\m{s}^n(\m{a}))$ reduces in at least $n^d$ steps.

\section{Predicative Interpretations}\label{s:pint}

In the following, let $\RS$ denote a constructor TRS 
that is compatible with $\gspop$. 
To simplify matters, we suppose for 
now that $\RS$ is also \emph{completely defined}.
Consider a rewrite rule 
${f(\sn{\vec{u}}{\vec{v}}) \to r} \in \RS$
that triggers an innermost rewrite step 
$$
s = C[f(\sn{\vec{u}\sigma}{\vec{v}\sigma})] \irew C[r\sigma] = t\tpkt
$$
Since normal forms and values coincide,
the rule is only triggered
if all arguments $\vec{u}\sigma,\vec{v}\sigma$  of the redex
are values.
Due to the limitations imposed by $\cref{gspop}{ia}$
and $\cref{gspop}{ts}$, it is not difficult 
to see that if $r\sigma$ is not a value itself, 
then at least all normal arguments are values.
We capture this observation in the set $\Tn$, 
defined as the least extension of 
values closed under $\FS$ operations containing
only values at normal argument positions:
$\Tn$ is the least set such that
(i) $\Val \subseteq \Tn$, and
(ii) if $f \in \FS$, $\vec{s} \subseteq \Val$ 
and $\vec{t} \subseteq \Tn$ 
then $f(\sn{\vec{s}}{\vec{t}}) \in \Tn$.
This set is closed under rewriting. 
\begin{lemma}\label{l:spopstar:tnderiv}
  Let $\RS$ be a completely defined TRS compatible with $\gspop$.
  If $s \in \Tn$ and $s \irew t$ then $t \in \Tn$.
\end{lemma}
\begin{IEEEproof}
  The Lemma follows by a straight forward inductive argument on Definition~\ref{d:gspop}.
\end{IEEEproof}
Since $\Tn$ contains in particular all basic terms, 
it follows that the runtime complexity function $\rci$ depends only on terms from $\Tn$. 

The \emph{predicative interpretation $\ints$} maps
terms from $\Tn$ to \emph{sequences} of \emph{normalised} terms
by separating normal from safe arguments.
Here a term is normalised if it is a term where arities of defined symbols 
correspond to the number of normal argument positions. 
We write $\fn$ for the symbol $f$ if it occurs in a normalised term.
To denote sequences of terms, 
we use a fresh variadic function symbol $\listsym$. 
Here variadic means that the arity of $\listsym$ 
is finite but otherwise arbitrary.  
We always write $\lseq{a}$ for $\listsym(\seq{a})$, 
in particular if we write $f(\seq{a})$ then $f \not = \listsym$.
We denote by $\LS[\FS]$ the set of \emph{sequences}
$\lseq{t}$ of normalised terms $\seq{t}$.
Abusing set-notation, we denote by $s \in \lseq{s}$ that $s=s_i$ for some $i \in \{1,\dots,n\}$.

The predicative interpretation $\ints$ is defined on $\Tn$ as follows:
If $t$ is a value, then $\ints(t) \defsym \nil$. Otherwise
if $t = f(\pseq{t})$, then
\begin{equation*}
  \ints(t) \defsym \lst{\fn(t_1, \dots, t_k)} \append \ints(t_{k+1}) \append \cdots \append \ints(t_{k+l}) \tpkt
\end{equation*}
Here the \emph{concatenation} operator $\append$ is defined on 
sequences such that $\lseq{s} \append \lseq{t} \defsym \lst{s_1~\cdots~s_n~t_1~\cdots~t_m}$.
We extend concatenation to terms by identifying terms $t$
with the singleton sequences $\lst{t}$, for instance $s \append t= \lst{s~t}$. 

In Fig.~\ref{fig:pi} we exemplify the predicative interpretation $\ints$
on a rewrite step of the $\RSsquare$ depicted in
Fig.~\ref{fig:square}. 
\begin{figure}
  \centering
  \begin{tikzpicture}
      \node (s) at (-2,0) {$\mTimes(\sn{\mS(\mS(\mZ)),\mS(\mZ)}{})$};
      \node (arr) at (0,0) {$\irew[\RSsquare]$};
      \node (t) at (2.5,0) {$\mPlus(\sn{\mS(\mZ)}{\mTimes(\sn{\mS(\mZ),\mS(\mZ)})})$};
      
      \node (is)[below of=s, yshift=-3mm] {$\lst{\fsn{\mTimes}(\mS(\mS(\mZ)),\mS(\mZ))}$};
      \node (it)[below of=t, yshift=-3mm] {$\lst{\fsn{\mPlus}(\mS(\mZ))~~\fsn{\mTimes}(\mS(\mZ),\mS(\mZ))}$};

      \draw[->] (s) to node[right] {$\ints(\cdot)$} (is);
      \draw[->] (t) to node[right] {$\ints(\cdot)$} (it);
    \end{tikzpicture}
  \caption{Predicative interpretation}
  \label{fig:pi}
\end{figure}


\section{The Small Polynomial Path Order on Sequences}\label{s:spop}

We define the \emph{small polynomial path order on sequences} $\LS[\FS]$.
As these serve a purely technical
reason, it suffices to represent the order via finite approximations $\gspopv[k]$.
The parameter $k \in \N$ controls the width of terms and sequences. 
We lift terms equivalence to sequences by disregarding order of elements:
$\lseq{s} \eqi \lseq{t}$ if
$s_i \eqi t_{\pi(i)}$ for all $i=1,\dots,n$
and some permutation $\pi$ on $\{1,\dots,n\}$.

\begin{definition} \label{d:gspopv} 
  Let $k \geqslant 1$, and let $\qp$ denote an admissible precedence.
  We define ${\gspopv[k]}$ inductively such that:
  \begin{enumerate}
  \item \label{d:gspopv:ia} 
    $f(\seq{s}) \gspopv[k] g(\seq[m]{t})$ if
    $f$ is a defined symbol,
    $g$ is below $f$ in the precedence
    and the following conditions hold:
    \begin{enumerate}
    \item all arguments $t_j$ are equivalent to proper subterms of $f(\seq{s})$; 
    \item $m \leqslant k$.
    \end{enumerate}
  \item \label{d:gspopv:ts}
    $f(\seq{s}) \gspopv[k] g(\seq[n]{t})$ if $f$ is recursive 
    and equivalent to $g$ in the precedence 
    and the following conditions hold:
    \begin{enumerate}
    \item $\tuple{s_1,\dots,s_n} \esupertermstrict \tuple{t_{\pi(1)},\dots,t_{\pi(n)}}$ for some permutation $\pi$;
    \item $n \leqslant k$.
    \end{enumerate}
  \item \label{d:gspopv:ialst} 
    $f(\seq{s}) \gspopv[k] \lseq[m]{t}$ and the following conditions hold:
    \begin{enumerate}  
    \item $f(\seq{s}) \gspopv[k] t_{j}$ for all $j = 1,\dots,m$;
    \item at most one element $t_j$ ($j \in \{1,\dots,m\}$) 
      contains defined symbols not below $f$ in the precedence;
    \item $m \leqslant k$.
    \end{enumerate}
  \item \label{d:gspopv:ms} 
    $\lseq{s} \gspopv[k] b$ where $b$ is equivalent to $b_1 \append \cdots \append b_n$ 
    and the following conditions hold:
    \begin{enumerate}
    \item $s_i \geqspopv[k] b_i$ for all $i = 1, \dots, n$;
    \item $s_{i_0} \gspopv[k] b_{i_0}$ for at least one $i_0 \in \{1, \dots, n\}$.
    \end{enumerate}
  \end{enumerate}
  Here $a \geqspopv[k] b$ denotes that either $a$ and $b$ are equivalent
  or $a \gspopv[k] b$ holds, and 
  $\tuple{s_1,\dots,s_k} \esupertermstrict \tuple{t_i,\dots,t_k}$
  denotes that the term $t_i$ is equivalent to a subterm of $s_i$ for all $i = 1,\dots,n$, and
  at least one $t_{i_0}$ is equivalent to a proper subterm of $s_{i_0}$ ($i_0 \in \{1,\dots,n\}$).
\end{definition}

In Fig.~\ref{fig:pi2} we demonstrate that 
predicative interpretation $\ints$ as exemplified in 
Fig.~\ref{fig:pi} embed the corresponding rewrite step 
into $\gspopv[2]$. Here we abbreviate 
$s = \fsn{\mTimes}(\mS(\mS(\mZ)),\mS(\mZ))$.

\begin{figure}
\centering
\begin{alignat*}{4}
  & \text{\lbl{1:}} &~~&& s 
      & \gspopv[2] \fsn{\mPlus}(\mS(\mZ)) 
      & \text{by \cref{gspopv}{ia}}\\
  & \text{\lbl{2:}} &&& s 
      & \gspopv[2] \fsn{\mTimes}(\mS(\mZ),\mS(\mZ))  
      & \text{by \cref{gspopv}{ts}}\\
  & \text{\lbl{3:}} &&&  s 
      & \gspopv[2] \lst{\fsn{\mPlus}(\mS(\mZ))~\fsn{\mTimes}(\mS(\mZ),\mS(\mZ))} 
      \quad & \text{by \cref{gspopv}{ialst}, using \lbl{1} and \lbl{2}}\\
  & \text{\lbl{4:}} &&& \lst{s} 
      & \gspopv[2] \lst{\fsn{\mPlus}(\mS(\mZ))~\fsn{\mTimes}(\mS(\mZ),\mS(\mZ))} 
      & \text{by \cref{gspopv}{ms}, using \lbl{3}}
\end{alignat*}  
\caption{Predicative interpretation revisited}
\label{fig:pi2}
\end{figure}

The orders $\gspopv[k]$ is defined so that following conditions are satisfied:
\begin{lemma} \label{l:approx}
  Let $k \geqslant 1$. We have
  \begin{enumerate}
  \item \label{l:approx:1} ${\gspopv[l]} \subseteq {\gspopv[k]}$ for all $l \leqslant k$,
  \item \label{l:approx:2} ${\eqi} \cdot {\gpopv_k} \cdot {\eqi} \subseteq {\gpopv_k}$, and
  \item \label{l:approx:3} if $a \gspopv[k] b$ then ${a \appendp c} \gspopv[k] {b \appendp c}$.
  \end{enumerate}
\end{lemma}
\begin{IEEEproof}
We focus on Property~\ref{l:approx:3}, the other
facts either follow by definition, or by a straight forward inductive argument.
In proof of Property~\ref{l:approx:3}
we perform case analysis on the last rule that concludes $a \gspopv[k] b$.

Suppose $a \cref{gspopv}{ia} b$ or $a \cref{gspopv}{ts} b$.
Then $a = f(\seq{s})$ and $b = g(\seq[m]{t})$.
Hence $a \append c = \lst{f(\seq{s})~u_1~\cdots~u_l}$ 
$b \append c = g(\seq[m]{t}) \append u_1 \append \cdots \append u_l$, 
where either $c$ is the sequence $\lseq[l]{u}$ or $c$ is the 
term $u_1$ and $l = 1$.
We conclude the lemma using \cref{gspopv}{ms}, employing the inequalities
$a \gspopv[k] b$ and $u_i \geqspopv[k] u_i$ for all $i \in \{1,\dots,l\}$.

Suppose $a \cref{gspopv}{ialst} b$.
Then $a = f(\seq{s})$ and $b = \lseq[m]{t}$.
We conclude $a \append c \gspopv[k] b \append c$ similar to above.

Suppose $a \cref{gspopv}{ms} b$.  
In this case $a = \lseq[n]{s}$ and
$b$ is equivalent to $b_1 \append \cdots \append b_n$ such that
$s_i \geqspopv[k] b_i$ for all $i = 1,\dots,n$ and 
$s_{i_0} \gspopv[k] b_{i_0}$ for at least one $i_0 \in \{1, \dots, n \}$.
Then 
$
a \append c = 
\lst{s_1~\cdots~s_n~u_1~\cdots~u_l}
$
and 
$b \append c$ is equivalent
to 
$b_1 \append \cdots \append b_n \append u_1 \append \cdots \append u_l$.
One application of \cref{gspopv}{ms} proves the lemma.
\end{IEEEproof}

The length of $\gspopv[k]$ descending sequences is expressed by the function
$\SSlow[k]$, given by
\begin{equation*}
  \SSlow[k](a) \defsym 1 + \max \{ \SSlow[k](b) \mid a \gspopv[k] b \}\tpkt
\end{equation*}
Here $a$ ranges over (normalised) terms and sequences.
As intermediate result we obtain that sequences act purely as containers with
respect to $\SSlow[k]$.
\begin{lemma} \label{l:slowsum} 
  Let $a = \lseq{t}$ be sequence. Then $\SSlow[k](a) = \sum_{i=1}^n \SSlow[k](t_i)$.
\end{lemma}
\begin{IEEEproof}
  Let $a = \lseq{s}$ be a sequence and 
  observe $\SSlow(a_1 \append a_2) \geqslant \SSlow(a_1) + \Slow(a_2)$.
  This is a consequence of Lemma~\eref{l:approx}{3}.
  Hence in particular 
  $
  \SSlow(a) =  \SSlow(s_1 \append \cdots \append s_n) \geqslant \sum_{i=1}^n \SSlow(s_i)
  $ 
  follows.

  To prove the inverse direction, 
  we show that $a \gspopv[k] b$ implies $\SSlow(b) < \sum_{i=1}^n \SSlow(s_i)$
  by induction on $\SSlow(a)$.
  The base case $\Slow(a) = 0$ follows trivially as the assumption $a \gspopv[k] b$
  is not satisfied.
  For the inductive step, observe that $a \gspopv[k] b$ follows due to \cref{gspopv}{ms}.
  Hence $b$ is equivalent to $b_1 \append \cdots \append b_n$ where $s_i \geqspopv[k] b_i$ for all 
  $i = 1,\dots,n$ and
  $s_{i_0} \gspopv[k] b_{i_0}$ for at least one $i_0 \in \{1,\dots,n\}$.
  In particular, $\SSlow(b_i) \leqslant \Slow[k](s_i)$ and $\Slow(b_{i_0}) < \Slow[k](s_{i_0})$.
  As we have $\SSlow(b_i) \leqslant \SSlow[k](b) < \SSlow(a)$ for all $i \in \{1,\dots,n\}$,
  induction hypothesis is applicable to $b$ and all $b_i$ ($i \in \{1,\dots,n\}$).
  It follows that 
  \begin{align*}
    \SSlow(b) = \sum_{t \in b} \SSlow(t) = \sum_{i=1}^n \sum_{t \in b_i} t = \sum_{i=1}^n \SSlow(b_i) < \sum_{i=1}^n \Slow[k](s_i) \tpkt
  \end{align*}
\end{IEEEproof}

Let $r$ and $d$ be natural number. We recursively define:
  \begin{equation*}
    \mc(r,d) \defsym
    \begin{cases}
      1 & \text{ if $r = 1$, and } \\
      \mc(r-1,d) \cdot k^{d+1} + 1 & \text{ otherwise} 
      \tpkt
    \end{cases}
  \end{equation*}
Below the argument $r$ will be instantiated by the rank 
and $d$ by the depth of recursion of a function symbol $f$.

The next lemma is a technical lemma to ease
the presentation of the proof of Theorem~\ref{t:spop}.
\begin{lemma}
\label{l:spop:1}
\mbox{}

\begin{enumerate}
\item Suppose $f(\seq{s}) \gspopv[k] g(\seq[m]{t})$ such
that $g$ is below $f$ in the precedence. 
Further suppose
\begin{equation*}
  \SSlow(g(\seq[m]{t})) \leqslant 
  \mc(\rk(g),\recdepth(g)) \cdot (2 + \sum_{i=1}^m \depth(t_i))^{\recdepth(g)}  
  \tpkt
\end{equation*}
Then also 
$\SSlow(g(\seq[m]{t})) \leqslant \mc(r-1,d) \cdot k^{d} \cdot (2 + u)^{\recdepth(g)}$
where $u \defsym \sum_{i=1}^n \depth(s_i)$, $r \defsym \rk(f)$ and $d \defsym \recdepth(f)$.
\item Suppose $f(\seq{s}) \gspopv[k] g(\seq[m]{t})$ such 
that $f$ is equivalent to $g$ in the precedence.
Further suppose that $\sum_{i=1}^m \depth(t_i) < \sum_{i=1}^n \depth(s_i)$
implies
\begin{equation*}
  \SSlow(g(\seq[m]{t})) \leqslant 
  \mc(\rk(g),\recdepth(g)) \cdot (2 + \sum_{i=1}^m \depth(t_i))^{\recdepth(g)}
\end{equation*}
Then also $\SSlow(g(\seq[m]{t})) \leqslant \mc(r,d) \cdot (1 + u)^{d}$
where $u \defsym \sum_{i=1}^n \depth(s_i)$, $r \defsym \rk(f)$ and $d \defsym \recdepth(f)$.
\end{enumerate}
\end{lemma}
\begin{IEEEproof}
We consider the first point of the proposition.
By assumption $f(\seq{s}) \cref{gspopv}{ia} g(\seq[m]{t})$.
Consider a term $t_j$ where $j \in \{1,\dots,m\}$. 
By definition, $t_j$ is equivalent to a proper subterms of $f(\seq{s})$
and it follows that $\depth(t_j) \leqslant \sum_{i=1}^n \depth(s_i) \symdef u$. 
Hence $\sum_{i=1}^m \depth(t_i) \leqslant k \cdot u$
since $m \leqslant k$ by definition. 
As $g$ is below $f$ in the precedence, we have 
$\rk(g) < r$ and $\recdepth(g) \leqslant d$. 
We conclude by monotonicity of $\mc$ and assumption
\begin{align*}
  \SSlow(g(t_1,\dots,t_m)) & \leqslant \mc(\rk(g),\recdepth(g)) \cdot (2 + k \cdot u)^{\recdepth(g)}
  \\
  & \leqslant \mc(r-1,d) \cdot k^{d} \cdot (2 + u)^{\recdepth(g)}
  \tpkt
\end{align*}

For the second point of the proposition, 
observe that by assumption $f(\seq{s}) \cref{gspopv}{ts}  g(\seq[m]{t})$.
Hence $m = n$ and the 
order constraints on arguments give
$\sum_{i=1}^n \depth(t_i) < \sum_{i=1}^n \depth(s_i) \symdef u$.  
Using $\rk(g) = r$ and $\recdepth(g) = d$, the
assumptions yield:
\begin{align*}
  \SSlow[k](g(\seq[n]{t})) 
    & \leqslant \mc(\rk(g),\recdepth(g)) \cdot (2 + \sum_{i=1}^m \depth(t_i))^{\recdepth(g)}
    \\
  & \leqslant \mc(r,d) \cdot (1 + u)^{d} \tpkt
\end{align*}
\end{IEEEproof}

\begin{theorem}\label{t:spop}
  Let $f$ be a defined symbol of recursion depth $d$.
  Then $\SSlow(f(\seq[n]{s})) \leqslant c \cdot {\bigl(\sum_{i=1}^n \depth(s_i)\bigr)}^{d}$
  for all values $\seq[n]{s}$.
  Here the constant $c \in\N$ depends only on $f$ and $k$.
\end{theorem}
\begin{IEEEproof}
Let $k$ be fixed.
To show the theorem, we show for all terms $f(\seq[n]{s})$,
whose arguments are constructor terms, that $f(\seq[n]{s}) \gspopv[k] b$
implies:
\begin{equation*}
    \SSlow(b) < \mc(\rk(f),\recdepth(f)) \cdot 
    \bigl(2 + \sum_{i=1}^n \depth(s_i)\bigr)^{\recdepth(f)} \tpkt
\end{equation*}
In proof we employ induction on $\rk(f)$ and side induction on 
$\sum_{i=1}^n \depth(s_i)$. 

Consider $g(\seq[m]{t})$ with $f \qp g$.
We state the induction hypothesis (IH) and 
side induction hypothesis (SIH).
IH states that if $g$ is below $f$ in the precedence then
\begin{equation*}
  \SSlow(g(\seq[m]{t})) \leqslant \mc(\rk(g),\recdepth(g)) \cdot (2 + \sum_{i=1}^m \depth(t_i))^{\recdepth(g)} \tkom
\end{equation*}
while SIH states that if $f$ and $g$ are equivalent in the precedence but
$\sum_{i=1}^m \depth(t_i) < \sum_{i=1}^n \depth(s_i)$ then
\begin{equation*}
  \SSlow(g(\seq[m]{t})) \leqslant \mc(\rk(g),\recdepth(g)) \cdot (2 + \sum_{i=1}^m \depth(t_i))^{\recdepth(g)} \tpkt  
\end{equation*}
%
%
Set $u \defsym \sum_{i=1}^n \depth(s_i)$, $r \defsym \rk(f)$ and $d \defsym \recdepth(f)$
and assume $f(\seq[n]{s}) \gspopv[k] b$.
We prove $\SSlow(b) < \mc(r,d) \cdot {(2 + u)}^{d}$.

In the base case $r = 1$ of the main induction, 
either $f(\seq[n]{s}) \cref{gspopv}{ts} b$ or $f(\seq[n]{s}) \cref{gspopv}{ialst} b$
as $f$ is minimal in the precedence.
For the base case $u = 0$ of the side induction
we see that $b=\nil$, hence $\SSlow[k](b) = 0$ and the theorem follows.
For the inductive step of the side induction let $u > 0$.
If $f$ is compositional, the assumptions give
$f(\seq[n]{s}) \cref{gspopv}{ialst} b$, hence 
$b$ is a sequence, in particular $b = \nil$ and the 
theorem follows. 
For the remaining case that $f$ is recursive,
we consider two sub-cases: 
(i) $b$ is a term, and
(ii) $b$ is a sequence.
In the former sub-case, $f(\seq[n]{s}) \cref{gspopv}{ts} b$
where $b = g(\seq[n]{t})$ and $g$ is 
equivalent to $f$ in the precedence.
The order constraints on arguments give 
$\sum_{i=1}^n \depth(t_i) < u$.
Employing $\recdepth(g) = d$ and $\rk(g) = r$, 
we conclude the sub-case by SIH:
\begin{align*}
  \SSlow(b) & \leqslant \mc(r,d) \cdot (2 + \sum_{i=1}^n \depth(t_i))^{d}
  < \mc(r,d) \cdot (2 + u)^{d} \tpkt
\end{align*}

In the second sub-case $b$ is a sequence $\lseq[m]{t}$
where $f(\seq[n]{s}) \cref{gspopv}{ialst} b$.
%
%
In particular, minimality of $f$ in the precedence gives $m \leqslant 1$.
Then the theorem follows trivially for $m = 0$, for $m = 1$ we conclude 
by the sub-case (i), additionally employing Lemma~\ref{l:slowsum}.

Now, we consider the step case of the main induction, let $r > 2$.
If $b = g(\seq[m]{t})$, then obviously $f \qp g$ and we conclude 
by Lemma~\ref{l:spop:1}. Note
that IH (SIH) yield the assumptions of the lemmas.

On the other hand suppose $b = \lseq[m]{t}$ and thus 
$f(\seq{a}) \gspopv[k] b$ follows by \cref{gspopv}{ialst}.
Then $m \leqslant k$ and $f(\seq{a}) \gspopv[k] t_{j}$ for all $j = 1,\dots,m$. 
Additionally at most one $t_{j_0}$ ($j_0 \in \{1,\dots,m\}$)
contains defined symbols not below $f$ in the precedence.
We analyse two sub-cases: either (i) $f$ is recursive or
(ii) $f$ is compositional. 
We consider the first sub-case. In this case, $f \sp g$ implies 
$d > \recdepth(g) \geqslant 0$.
Lemma~\ref{l:spop:1} yields:
\begin{align*}
  \SSlow(t_j) & \leqslant \mc(r-1,d) \cdot k^{d} \cdot (2 + u)^{d-1} 
  \tag{\dag}
  \\
  \SSlow(t_{j_0}) & \leqslant \mc(r-1,d) \cdot k^{d} \cdot (2 + u)^{d-1} + \mc(r,d) \cdot (1 + u)^{d} \tpkt
\end{align*}
Here~$(\dag)$ holds if $j \not = j_0$.
Recall $m \leqslant k$ and $\mc(r-1,d) \cdot k^{d+1} < \mc(r,d)$.
We conclude with Lemma~\ref{l:slowsum}:
\begin{align*}
  \SSlow(b) & = \SSlow(t_{j_0}) + \sum_{j=1,j\not=j_0}^m \SSlow(t_j) \\
  & \leqslant k \cdot ( \mc(r-1,d) \cdot k^{d} \cdot (2 + u)^{d-1} ) + \mc(r,d) \cdot (1 + u)^{d} \\
  & < \mc(r,d) \cdot (2 + u)^{d-1} + \mc(r,d) \cdot (1 + u)^{d} \\
  & \leqslant \mc(r,d) \cdot (2 + u)^{d}\tpkt
\end{align*}

On the other hand, consider the second sub-case that $f$ is compositional.
Conclusively $f(\seq{a}) \gspopv[k] t_{j}$ can be strengthened to 
$f(\seq{a}) \cref{gspopv}{ia} t_{j}$.
Employing that $d \geqslant \recdepth(g)$ for all symbols $g$ 
below $f$ in the precedence, 
using Lemma~\ref{l:slowsum} and Lemma~\ref{l:spop:1} we see
\begin{align*}
  \SSlow(b)  = \sum_{j=1}^m \SSlow(b_j) 
  & \leqslant k \cdot (\mc(r-1,d) \cdot k^{d} \cdot (2 + u)^{d})
  \\
  & < \mc(r,d) \cdot (2 + u)^{d} \tpkt
\end{align*}
The theorem follows.
\end{IEEEproof}

\section{Predicative Embedding}\label{s:embed}

In this section we prove that for some 
constant $k \in\N$ depending only on the considered TRS $\RS$,
each innermost rewrite step gives rise 
to a $\gspopv[k]$ descent under predicative interpretation $\ints$.
To ease the presentation, we provide the following auxiliary lemma.

\begin{lemma}\label{l:embed:aux}
  Let $s = f(\sn{\seq[l]{s}}{\vec{s}})$ be a basic term, let 
  $t$ be a term of size up to $\ell \in \N$
  and let $\sigma$ be a substitution that maps variables to values.
  If $s \gspop t$ then 
  (i) $\fn(s_1\sigma, \dots, s_l\sigma) \gspopv[\ell] u$ for all $u \in \ints(t\sigma)$ and further, 
  (ii) at most one $u \in \ints(t\sigma)$ contains a defined symbols not below $f$ in the precedence.
\end{lemma}
\begin{IEEEproof}
  We prove Property (i) by induction on $\gspop$
  and case analysis on the last rule concluding $s \gspop t$.
  Consider the case $s \cref{gspop}{st} t$  where
  $s_i \geqspop t$ holds for some argument $s_i$ of $s$.
  Since only case \cref{gspop}{st} applies on values, 
  we see that $t$ is equivalent to subterm of $s_i$, in 
  particular $t\sigma$ is a value and $\ints(t\sigma) = \nil$.
  Trivially the lemma follows.

  Next, consider the case $s \cref{gspop}{ia} t$. 
  Then $t = g(\pseq[k][n]{t})$.
  Abbreviate $a = \ints(t_{k+1}\sigma) \append \cdots \append \ints(t_{k+n}\sigma)$ and 
  hence by definition $\ints(t\sigma) = \lst{\gn(t_1\sigma, \dots, t_k\sigma)} \append a$.
  As $s \gspop t_j$ for all safe arguments $t_j$ of $t$, 
  induction hypothesis gives
  $\fn(s_1\sigma, \dots, s_l\sigma) \gspopv[k] u$ for all $u \in a$.
  We verify 
  \begin{equation*}
    \label{eq:embed}
    \fn(s_1\sigma, \dots, s_l\sigma) 
    \gspopv[\ell] \gn(t_1\sigma, \dots, t_k\sigma) 
    \tag{\ddag} \tpkt
  \end{equation*}
  By the ordering constraints imposed by \cref{gspop}{ia}, 
  the defined symbol $g$ is below the defined symbol $f$ in the precedence, 
  and $s \nsubtermstrict t_j$ for all normal arguments $t_j$ of $t$.
  The latter reveals that
  the instances $t_j\sigma$ are equivalent to proper subterms of 
  the normalised term $\fn(s_1\sigma,\dots,s_l\sigma)$.
  As trivially $k$ is bounded by the size of $t$,
  one application of \cref{gspopv}{ia} concludes $(\ddag)$.

  Finally, consider the case $s \cref{gspop}{ts} t$.
  Then $t = g(\pseq[l][n]{t})$ where $g$ is equivalent to $f$ in the precedence.
  By reasoning similar to the case \cref{gspop}{st},
  the ordering constraint $\tuple{s_1,\dots,s_k} \gspop \tuple{t_{\pi(1)},\dots,t_{\pi(k)}}$
  on normal arguments reveals
  $\tuple{s_1\sigma,\dots,s_k\sigma} \esupertermstrict \tuple{t_{\pi(1)}\sigma,\dots,t_{\pi(k)}\sigma}$.
  As $l$ is trivially bounded by the size of $t$ we conclude \eqref{eq:embed} by \cref{gspopv}{ts}.  
  Using the ordering constraints on safe arguments of $t$, 
  we see that all safe arguments $t_j$ of $t$ are values, 
  in particular $\ints(t_j\sigma) = \nil$. 
  The lemma follows.

  For Property (ii) a straight forward induction reveals that $t$ contains at most 
  one defined symbol not below $f$ in the precedence. Here we make essential use 
  of Condition (iii) in \cref{gspop}{ia}.
  Then we conclude by the shape of $\sigma$ and definition of predicative interpretations.
\end{IEEEproof}

\begin{lemma}\label{l:embed}
  Let $\RS$ be a completely defined TRS compatible with $\gspop$. Let 
  denote the maximal size of a right-hand side in $\RS$.
  If $s \in \Tn$ and $s \irew t$ then $\ints(s) \gspopv[\ell] \ints(t)$.
\end{lemma}
\begin{IEEEproof}
  Let $s \in \Tn$ and consider an innermost rewrite 
  step $s \irew t$. 
  We prove the lemma by induction on the rewrite context.
  In the base case, the context is empty, i.e.,
  $s = l\sigma$ and $t = r\sigma$ 
  for some rule ${l \to r} \in \RS$
  where $l = f(\sn{\seq[m]{l}}{\vec{l}})$ is a basic term.
  Since $\RS$ is a completely defined TRS, 
  all arguments of $l\sigma$ are values.
  Further compatibility gives $l \gspop r$, and
  hence all preconditions of Lemma~\ref{l:embed:aux} met.
  We conclude $\fn(l_1\sigma, \dots, l_m\sigma) \gspopv[\ell] u$ 
  for all terms $u \in \ints(r\sigma)$, 
  where at most one $u$ is not constructed from 
  defined symbols below $\fn$ in the precedence.
  Exploiting that $\ints$ erases values, hence in particular
  images of $\sigma$, it is not difficult to prove that the length 
  of the sequence $\ints(r\sigma)$ is bounded by $\size{r} \leqslant \ell$.
  In total we obtain 
  $
  \ints(s) = \lst{\fn(l_1\sigma, \dots, l_m\sigma)}
  \gspopv[\ell] \ints(t)
  $
  by one application of \cref{gspopv}{ialst} followed by one 
  application of \cref{gspopv}{ms}.

  Consider now the stepping case
  $s = f(\sn{\vec{v}}{s_1,\dots,s_i,\dots,s_n})$
  $t = f(\sn{\vec{v}}{s_1,\dots,t_i,\dots,s_n})$
  and $s_i \irew t_i$. Here we use that since $s \in \Tb$
  all normal arguments $\vec{v}$ are values and cannot be rewritten.
  By induction hypothesis $\ints(s_i) \gspopv[\ell] \ints(t_i)$.
  Using Lemma \ref{l:approx} we see
  \begin{align*}
   \ints(s\sigma) & = 
   \fn(\vec{v}) \append \ints(s_{1}) \append \cdots \append \ints(s_i) \append \cdots \append \ints(s_{n})\\
   & \gspopv[\ell]
  \fn(\vec{v}) \append \ints(s_{1}) \append \cdots \append \ints(t_i) \append \cdots \append \ints(s_{n}) = \ints(t\sigma) \tpkt
  \end{align*}
\end{IEEEproof}

Putting things together, we arrive at the proof of the main theorem.
\begin{IEEEproof}[Proof of Theorem~\ref{t:spopstar}]
  Let $\RS$ denote a predicative recursive TRS of degree $d$. 
  We prove the existence of constant $c \in \N$ such that for all 
  values $\vec{u},\vec{v}$, the derivation height of the start term 
  $f(\svec{u}{v})$ with respect to $\irew$
  is bounded by $c \cdot n^k$ where $n$ is the sum of the 
  depths of normal arguments $\vec{u}$.
  
  Without loss of generality we can assume that $\RS$ is completely
  defined. Otherwise we add sufficient rules to 
  $\RS$ that make it completely defined. 
  For this we suppose, without loss of generality, 
  that the signature $\FS$ contains a constructor $\bot$ such that $f \qp \bot$ for all $f \in \DS$.
  Note that if we add such a symbol, then still the precedence 
  underlying $\gspop$ is admissible and the depth of recursion $d$ does not increase.
  Call a normal form $s$ \emph{garbage} the root of $s$ is defined.
  We extend $\RS$ by adding the \emph{garbage rules} rules $s \to \bot$ 
  for all normal forms $s$ which are garbage.
  Clearly $s \gspop \bot$ by one application of \cref{gspop}{ia},
  conclusively $\RSbot$ is compatible with $\gspop$ too.
  Denote by $s_\bot$ the result of replacing garbage in a term $s$ by $\bot$, that is
  $s_\bot$ is the unique normal form of $s$ with respect to the garbage rules.
  Since garbage rules do not overlap with the constructor TRS $\RS$,
  a straight forward inductive argument reveals that if $s \irew[\RS] t$ then 
  also $s_\bot \irst[\RSbot] t_\bot$. 
  Conclusively it suffices to estimate the number of rewrite steps induced by 
  the completely defined TRS $\RSbot$.

  Set $\ell$ to denote the maximal size of a right-hand side of a rule in $\RSbot$, 
  and observe that $\ell$ is well defined.
  Consider a maximal derivation
  $f(\svec{u}{v}) \irew[\RSbot] t_1 \irew[\RSbot] \cdots \irew[\RSbot] t_{n}$.
  Let $i \in \{0, \dots, n-1\}$.
  By Lemma~\ref{l:spopstar:tnderiv} it follow that $t_i \in \Tn$, 
  and consequently
  $\ints(t_i) \gspopv[\ell] \ints(t_{i+i})$ due to Lemma~\ref{l:embed}.
  So in particular the length $n$ is bounded by 
  the length of $\gspopv[\ell]$ descending sequences starting 
  from $\ints(f(\svec{u}{v})) = \lst{\fn(\vec{u})}$.
  Additionally using Lemma~\ref{l:slowsum},
  Theorem~\ref{t:spop} gives the constant $c \in \N$ as desired.
\end{IEEEproof}


\section{Completeness Results}\label{s:completeness}

In this section we show that the small polynomial path order \POPSTARS\
is complete. Indeed we can even show something stronger. 
Let $\RS$ be a TRS that makes only use of $d$ nestings of
safe recursion, then $\RS$ is predicative recursive of degree $d$.
Due to the weak form of safe composition we have the inclusion that
$\mathcal{B}_{\mathsf{wsc}} \subseteq \mathcal B$.
Concerning the converse inclusion, the following lemma states that 
the class $\mathcal{B}_{\mathsf{wsc}}$
is large enough to capture all the polytime computable functions.

\begin{lemma}
\label{lem:Bwsc}
Every polynomial time computable function belongs to 
$\bigcup_{k \in \mathbb N}\mathcal{B}^{k,0}_{\mathsf{wsc}}$.
\end{lemma}

One can show this fact by following the proof of Theorem 3.7 in
\cite{HW99}, where the unary variant of $\Bwsc$ is defined and the
inclusion corresponding to Lemma \ref{lem:Bwsc} is shown.
We give the proof of Lemma~\ref{lem:Bwsc} in detail below, but
first assume the lemma in order to succinctly state our
completeness result.

\begin{theorem}
\label{thm:strong_completeness}
For any $\Bwsc$-function $f$
there exists a confluent TRS
$\RS_f$ that is predicative recursive of degree $d$, where $d$
equals the maximal number of nested application of ($\m{SRN}$) in the
definition of $f$.
\end{theorem}

The completeness of \POPSTARS\ for the polytime computable functions is an
immediate consequence of Lemma \ref{lem:Bwsc} and Theorem
\ref{thm:strong_completeness}.
The witnessing TRS $\RS_f$ for $f \in \Bwsc$ in Theorem
\ref{thm:strong_completeness} is obtained via a term rewriting
characterisation of the class $\Bwsc$.
The term rewriting characterisation expresses
the definition of $\Bwsc$ as an \emph{infinite} TRS
$\RS_{\Bwsc}$ where the equations in Fig.~\ref{fig:Bwsc} are oriented
from left to right. 
Here binary words are formed from the constructor symbols $\varepsilon$, $\mS_0$
and $\mS_1$.

We define a one-to-one correspondence between the class $\Bwsc$ of
functions and the set of function symbols for $\RS_{\Bwsc}$ as follows.
The function symbols $\ZEROP{k}{l}, \PROJP{k}{l}{j}, \PREDP, \CONDP$
correspond respectively to the initial functions 
$O^{k,l}, I^{k,l}_j, P, C$ of $\Bwsc$. 
The symbol $\SUBP[h,\seq[k]{i},\vec{g}]$ is used to denote the
function obtained by composing functions $h$ and $\vec{g}$ according to
the schema of ($\m{WSC}$)
Finally, the function symbol
$\SRN [g, h_0, h_1]$ corresponds
to the function defined by safe recursion on notation from 
$g$, $h_0$ and $h_1$ in accordance to the schema ($\m{SRN}$).  
It is easy to see that 
$\RS_{\Bwsc}$ is a constructor TRS.
Further $\RS_{\Bwsc}$ is a orthogonal TRS, thus confluent.

\begin{IEEEproof}[Proof of Theorem \ref{thm:strong_completeness}]
Let $f$ be an arbitrary function from $\Bwsc$.
By induction according to the definition of $f$ in $\Bwsc$ we show the
existence of a TRS $\RS_f$ and a precedence $\qp_f$ such that
\begin{enumerate}
\item $\RS_f$ is a finite restriction of $\RS_{\Bwsc}$,
\label{comp_thm:1}
\item $\RS_f$ contains the rule(s) that defines the function symbol
 $\m{f}$ corresponding to $f$,
\label{comp_thm:2}
\item $\RS_{f}$ is compatible with the \POPSTARS induced by $\qp_{f}$, 
\label{comp_thm:3}
\item $\m{f}$ is maximal in the precedence $\qp_f$, and
\label{comp_thm:4}
\item  the maximal depth of recursion of the function symbols,
 i.e., $\recdepth (\m{f})$, equals the maximal number of nested
 application of ($\m{SRN}$) in the definition of $f$ in $\Bwsc$.
\label{comp_thm:5}
\end{enumerate}
To exemplify the construction we consider the step case 
that $f$ is defined from some functions 
$g, h_0, h_1 \in \Bwsc$ by the schema ($\m{SRN}$).
By induction hypothesis we can find witnessing TRSs
$\RS_{g}, \RS_{h_0}, \RS_{h_1}$ and witnessing precedences
$\qp_{g}, \qp_{h_0}, \qp_{h_1}$ respectively for
$g, h_0, h_1$.
Extend the set of function symbols by the recursive symbol
$\m{f} :\equiv \SRN [\m{g}, \m{h}_0, \m{h}_1]$.
Let $\RS_f$ be the TRS consisting of
$\RS_g$, $\RS_{h_0}$, $\RS_{h_1}$ and the following three rules:
\begin{enumerate}
\item 
$\m{f} (\varepsilon, \vec{x};\vec{y}) \to 
 \m{g} (\vec{x};\vec{y})$.
\label{SRN:1}
\item
$\m{f} (\mS_i (; z), \vec{x};\vec{y})
 \to \m{h}_i (z,\vec{x};\vec{y}, 
             \m{f} (z, \vec{x};\vec{y}))$
$( i = 0, 1)$.
\label{SRN:2}
\end{enumerate}
Define the precedence $\qp_f$ extending
$\qp_{g} \cup \qp_{h_0} \cup \qp_{h_1}$ by
\begin{itemize}
\item $\m{f} \ep \m{f}$ and
\item $\m{f} \sp \m{g}'$
for any $\m{g}' \in \{ \m{g}, \m{h}_0, \m{h}_1 \}$.
\end{itemize}
Let $\gspop$ be the \POPSTARS induced by $\qp_f$.
Then it is easy to check that $\RS_f$ enjoys Condition
 \ref{comp_thm:1}) and \ref{comp_thm:2}).
In order to show Condition~\ref{comp_thm:3}), it suffices to
orient the three new rules by $\gspop$.
For the rule in \ref{SRN:1}),
$\m{f} (\varepsilon, \vec{x};\vec{y}) 
 \cref{gspop}{ia} \m{g} (\vec{x};\vec{y})$
holds by the definition of $\qp_f$.
For the remaining two rules in \ref{SRN:2}) we only orient the case
$i = 0$.
It is clear that
$\m{f} (\mS_0 (; z), \vec{x};\vec{y})  
 \cref{gspop}{st} u$
holds for any $u$ from $z, \vec x, \vec y$.
In particular $\mS_0 (;z) \cref{gspop}{st} z$ holds.
Hence 
$\m{f} (\mS_0 (; z), \vec{x};\vec{y}) 
 \cref{gspop}{ts} 
 \m{f} (z, \vec{x};\vec{y})$
holds.
This together with the definition of the precedence $\qp_f$ allows us to conclude 
$$\m{f} (\mS_0 (; z), \vec{x};\vec{y}) 
 \cref{gspop}{ia} 
 \m{h}_0 (z, \vec x; \vec y, \m{f} (z, \vec{x};\vec{y})).$$
Consider Condition \ref{comp_thm:4}).
For each $g' \in \{ g, h_0, h_1 \}$, $\m{g}'$ is maximal in the precedence
$\qp_{g'}$ by induction hypothesis for $g'$.
Hence by the definition of $\qp_f$,
$\m{f}$ is maximal in $\qp_f$.
It remains to show Condition \ref{comp_thm:5}).
Since $\m{f}$ is a recursive symbol
$\recdepth (\m{f}) = 1+
 \max \{ \recdepth (\m{g}), \recdepth (\m{h}_0), \recdepth (\m{h}_1)
      \}$.
Without loss of generality let us suppose 
$\recdepth (\m{g}) =
 \max \{ \recdepth (\m{g}), \recdepth (\m{h}_0), \recdepth (\m{h}_1)
      \}$.
Then by induction hypothesis for $g$,
$\recdepth (\m{g})$ equals the maximal number of nested application of
($\m{SRN}$)in the definition of $g$ in $\Bwsc$.
Hence 
$\recdepth (\m{f}) = 1 + \recdepth (\m{g})$ 
equals the one in the definition of $f$ in $\Bwsc$.
\end{IEEEproof}

In the sequel of this section, we provide the (technical) proof of
Lemma~\ref{lem:Bwsc}.
In what follows we totally follow presentations by W.G. Handley and
S.S. Wainer in  
\cite[Section 3]{HW99}.
We start with defining the $k$-th iteration $f^{(k)} \in \Bwsc$ of
$f \in \Bwsc$ by
\begin{align*}
f^{(0)} (\vec x; \vec y, z) & = z, \text{ and} \\
f^{(k+1)} (\vec x; \vec y, z) & = f(\vec x; \vec y, f^{(k)} (\vec x; \vec y, z))\tpkt
\end{align*}
By the definition it is easy to see that
$f^{(k)} (\vec x; \vec y, f^{(l)} (\vec x; \vec y, z)) =
 f^{(k+l)} (\vec x; \vec y, z)$
holds.

\begin{lemma}
\label{lem:iter}
(Cf. \cite[Lemma 3.3]{HW99})
Let $p$ be an $n$-ary polynomial with non-negative coefficients.
If $f(x_1, \dots, x_n; \vec y, z) \in \Bwsc$, then there exists a
 function $\iter [p, f] \in \Bwsc$ such that 
$\iter [p,f] (\vec x; \vec y, z) =
 f^{(p(|\vec x|))} (\vec x; \vec y, z)$.
\end{lemma}

\begin{proof}
We define a witnessing function $\iter [p,f]$ by induction over the
 construction of the polynomial $p$.
In case that $p$ is a trivial polynomial, i.e.,
$p (\vec x) = 0$ or $p(\vec x) = 1$, the choice of
$\iter [p,f]$ is clear.

\textsc{Case.}
$p(\vec x) = q(\vec x) + r(\vec x)$ for some polynomials $q, r$:
In this case $\iter [p,f]$ is defined by 
$\iter [p,f] (\vec x; \vec y, z) =
 \iter [q,f] (\vec x; \vec y, \iter [r, f] (\vec x; \vec y, z))$.

\textsc{Case.}
$p(\vec x) = x_j \cdot q(\vec x)$ for some $j \in \{ 1, \dots n \}$ and
 for some polynomial $q(\vec x)$:
In this case we define an auxiliary function $\loopf [q, f] \in \Bwsc$ by
\begin{align*}
\loopf [q, f] (\epsilon, \vec x; \vec y, z) & =  z, \\
\loopf [q, f] (ui, \vec x; \vec y, z) & = 
\iter [q, f] (\vec x; \vec y, \loopf [q, f] (u, \vec x; \vec y, z))
& (i = 0, 1)
\end{align*}
Then $\loopf [p, f]$ is defined by 
$\iter [q, f] (\vec x; \vec y, z) =
 \loopf [q, f] (x_j, \vec x; \vec y, z)$.
One can check that 
$f^{(|u| \cdot q(|\vec x|))} (\vec x; \vec y, z) =
 \loopf [q, f] (u, \vec x; \vec y, z)$
holds by induction on $|u|$.
\end{proof}

We write $y \oplus x$ to denote the sequence $y$ followed by $|x|$
$0$'s. 
Namely the operator $\oplus$ satisfies the equations
$y \oplus \epsilon = y$ and
$y \oplus (x0) = y \oplus (x1) = (y \oplus x) 0$.
We will write $y \oplus x \oplus x'$ instead of
$(y \oplus x) \oplus x'$.

\begin{lemma}
\label{lem:oplus}
(Cf. \cite[Lemma 3.4]{HW99})
Let $f, h$ be arbitrary functions.
If $f$ and $h$ enjoy the condition
\begin{align*}
f(\vec x, \epsilon, \vec b, c, d) &= c, \text{ and}\\
f(\vec x, a0, \vec b, c, d) &=
f(\vec x, a1, \vec b, c, d) =
h(\vec x, d \oplus a, \vec b, f(\vec x, a, \vec b, c, d)),
\end{align*}
then $f$ also enjoys the condition
\begin{align*}
f(\vec x, u, \vec b, f(\vec x, t, \vec b, c, w), w \oplus t) =
 f(\vec x, t \oplus u, \vec b, c, w)\tpkt
\end{align*}
\end{lemma}

\begin{proof}
By induction on $|u|$.
For the base case
$f(\vec x, \epsilon, \vec b, f(\vec x, t, \vec b, c, w), w \oplus t) =
 f(\vec x, t, \vec b, c, w) =
 f(\vec x, t \oplus \epsilon, \vec b, c , w)$.
For the induction step
\begin{align*}
f(\vec x, ui, \vec b, f(\vec x, t, \vec b, c, w), w \oplus t)
&=
h(\vec x, w \oplus t \oplus u, \vec b, 
  f(\vec x, u, \vec b, f(\vec x, t, \vec b, c, w), w \oplus t)) \\
&=
h(\vec x, w \oplus t \oplus u, \vec b,
  f(\vec x, t \oplus u, \vec b, c, w))
& \text{(by IH)} \\
&=
f(\vec x, t \oplus ui, \vec b, c, w).
\end{align*}
\end{proof}

Let $p$ be a polynomial with non-negative coefficients.
Let us define two functions $\oplus_p$ and $\ominus_p$ by
$\oplus_p (\vec x; y) = \iter [p, S_0] (\vec x; y)$ and
$\ominus_p (\vec x; y) = \iter [p, P] (\vec x; y)$ for the predecessor
function $P$.
Then $\oplus_p, \ominus_p \in \Bwsc$ by Lemma \ref{lem:iter}.
By definition
$\oplus_p (\vec x; y)$ denotes the sequence $y$ followed by
$p(|\vec x|)$ $0$'s and $\ominus_p (\vec x; y)$ denotes the sequence
consisting of the first $|y| - p (|\vec x|)$ symbols of $y$
(if $p (|\vec x|) \leqslant |y|$).
In the following we use the operator ``$\min$'' in a modified sense that
$\min (x, y) = x$ if $|x| < |y|$, or otherwise $\min (x, y) = y$.

\begin{lemma}
\label{lem:f[p]}
(Cf. \cite[Lemma 3.5]{HW99})
Let $f$ and $h$ enjoy the condition in the premise of Lemma
 \ref{lem:oplus}.
Let $p$ be a polynomial $p$ with non-negative coefficients.
If there exists a function $h' \in \Bwsc$ such that
$h'(\vec x, a, \vec b, c) =
 h(\vec x, a, \vec b, c)$
holds for all $a, c$,
then there exists
 a function $f[p] \in \Bwsc$ such that
$f[p] (\vec x; a, \vec b, \vec c, d) =
 f(\vec x, \min (a, \oplus_p (\vec x; \epsilon)), \vec b, c, d)$
holds for all $a, c$ and $d$.
\end{lemma}

\begin{proof}
By induction over the construction of the polynomial $p$.
In the special case that $p (\vec x) =0$ the choice of the witnessing
 function $f[p]$ is clear.

\textsc{Case.}
$p(\vec x) = 1$:
In this case $f[p] \in \Bwsc$ is defined by
\begin{align*}
 f[p] (\vec x; a, \vec b, c, d) =
 \begin{cases}
   c & \text{if $a = \epsilon$,} \\
   h'(\vec x; d, \vec b, c) & \text{otherwise.}
 \end{cases}
\end{align*}
Then the function $f[p]$ is as required since by the modified definition of
$\min$, $\min (a, 0) = \epsilon$ if $a = \epsilon$, or otherwise
$\min (a, 0) = 0$. 

\textsc{Case.}
$p(\vec x) = q(\vec x) + r(\vec x)$ for some polynomials $q$ and $r$:
In this case a function $f[p]$ is defined by
\begin{align*}
 f[p] (\vec x; a, \vec b, c, d) =
 f[q] (\vec x; \ominus_r (\vec x; a), \vec b, 
       f[r] (\vec x; a, \vec b, c, d), \oplus_r (\vec x; d)) \tpkt 
\end{align*}
We show that 
$f[p] (\vec x; a, \vec b, c, d) =
 f(\vec x, \min (a, \oplus_p (\vec x; \epsilon)), \vec b, c, d)$
holds by (sub)case-analysis.

\textsc{Subcase 1.}
$|a| \leqslant r (|\vec x|)$:
In this case $\ominus_r (\vec x; a) = \epsilon$.
Hence the following equality holds:
\begin{align*}
  f[p] (\vec x; a, \vec b, c, d) 
  &=
  f[q] (\vec x; \epsilon, \vec b, 
  f[r] (\vec x; a, \vec b, c, d), \oplus_r (\vec x; d)) \\
  &=
  f(\vec x, \min (0, \oplus_q (\vec x; \epsilon)), \vec b, 
  f[r] (\vec x; a, \vec b, c, d), \oplus_r (\vec x; d)
  )
  & \text{(by IH on $q$)} \\
  &=
  f(\vec x, \epsilon, \vec b, 
  f[r] (\vec x; a, \vec b, c, d), \oplus_r (\vec x; d)
  ) \\
  &=
  f[r] (\vec x; a, \vec b, c, d)
  & \text{(by Lemma \ref{lem:oplus})} \\
  &=
  f(\vec x, \min (a, \oplus_r (\vec x; \epsilon), \vec b, c, d)
  & \text{(by IH on $r$)} \\
  &=
  f(\vec x, \min (a, \oplus_p (\vec x; \epsilon), \vec b, c, d)
\end{align*}
Here the last equality follows as    
$\min (a, \oplus_r (\vec x; \epsilon)) = a = \min (a, \oplus_p (\vec x; \epsilon))$.

\textsc{Subcase 2.}
$r (|\vec x|) < |a| \leqslant q (|\vec x|) + r (|\vec x|)$:
In this case
$|\ominus_r (\vec x; a)| = |a| - r (|\vec x|) > 0$ and further
$\min (\ominus_r (\vec x; a), \oplus_q (\vec x; \epsilon)) = 
 \ominus_r (\vec x; a)$
hold.
Hence the following equality holds:
\begin{align*}
f[q] (\vec x; a, \vec b, c, d)
&=
f[q] (\vec x; \ominus_r (\vec x; a), \vec b, 
      f[r] (\vec x; a, \vec b, c, d), \oplus_r (\vec x; d)) \\
&=
f(\vec x, \ominus_r (\vec x; a), \vec b, 
  f(\vec x, \oplus_r (\vec x; \epsilon), \vec b, c, d),
  \oplus_r (\vec x; d)
 )
& \text{(by IH on $q$ and $r$)} \\
&=
f(\vec  x, \oplus_r (\vec x; \epsilon) \oplus \ominus_r (\vec x; a), 
  \vec b, c, d)
& \text{(by Lemma \ref{lem:oplus})} \\
&=
f(\vec x; a, \vec b, c, d) \tpkt
\end{align*}
The last equality holds since 
$|\oplus_r (\vec x; \epsilon) \oplus \ominus_r (\vec x; a)| = |a|$ and
$f(\vec x, a, \vec b, c, d) = f(\vec x, a', \vec b, c, d)$ 
if $|a| = |a'|$ by the assumption on $f$.
This allows us to conclude
$$f[q] (\vec x; a, \vec b, c, d) =
 f(\vec x, \min (a, \oplus_p (\vec x; \epsilon)), \vec b, c, d)\tpkt$$

\textsc{Subcase 3.}
$q (|\vec x|) + r (|\vec x|) < |a|$:
In this case 
$|\ominus_r (\vec x; a)| = |a| - r (|\vec x|) > 0$ and further
$\min (\ominus_r (\vec x; a), \oplus_q (\vec x; \epsilon)) = 
 \ominus_q (\vec x; \epsilon)$.
Hence the following equality holds:
\begin{align*}
f[p] (\vec x; a, \vec b, c, d)
& =
f[q] (\vec x; \ominus_r (\vec x; a), \vec b, 
      f[r] (\vec x; a, \vec b, c, d), \oplus_r (\vec x; d)) \\
& =
f(\vec x, \oplus_q (\vec x; \epsilon), \vec b, 
  f(\vec x, \oplus_r (\vec x; \epsilon)), \oplus_r (\vec x; d)
 )
& \text{(by IH on $q$ and $r$)} \\
& =
f(\vec x, \oplus_q (\vec x; \epsilon), \vec b, c, d)
& \text{(by Lemma \ref{lem:oplus})} \\
& =
f(\vec x, \min (a, \oplus_p (\vec x; \epsilon)), \vec b, c, d).
\end{align*}
This completes the case.

\textsc{Case.}
$p (\vec x) = x_j \cdot q (\vec x)$ for some $j$ and for some polynomial
 $q$:
In this case by IH there exists a witnessing function 
$f[q] \in \Bwsc$ on $q$.
Let us define a polynomial $q'$ by 
$q' (z, \vec x) = q (\vec x)$.
Then a witnessing function $f[q'] \in \Bwsc$ can be defined by
$f[q'] (z, \vec x; a, \vec b, c, d) =
 f[q] (\vec x; a, \vec b, c, d)$
since $q' (z, \vec x) = q(\vec x)$ holds.
In order to define $f[p]$ we introduce an auxiliary polynomial $p'$ by 
$p' (z, \vec x) = z \cdot q' (z, \vec x)$.
Further we define an auxiliary function
$f^1_{p'} \in \Bwsc$ via $f[q']$ by
\begin{align*}
f^1_{p'} (\epsilon, \vec x; a, \vec b, c, d) &= c, \\
f^1_{p'} (zi, \vec x; a, \vec b, c, d) &=
f[q'] (z, \vec x; \ominus_{p'} (z, \vec x; a), \vec b,
       f^1_{p'} (z, \vec x; a, \vec b, c, d),
       \oplus_{p'} (z, \vec x; a)
      ).
& (i = 0, 1)
\end{align*}
Now a function $f[p] \in \Bwsc$ is defined by
$f[p] (\vec x; a, \vec b, c, d) =
 f^1_{p'} (x_j, \vec x; a, \vec b, c, d)$.

\begin{claim}
$f^1_{p'} (z, \vec x; a, \vec b, c, d) =
 f(\vec x; \min (a, \oplus_{p'} (z, \vec x; \epsilon)),
   \vec b, c, d)$.
\end{claim}

Assuming the claim, we can conclude that
\begin{align*}
 f[p] (\vec x; a, \vec b, c, d) 
 = & f^1_{p'} (x_j, \vec x; a, \vec b, c, d) \\
 = & f(\vec x, \min (a, \oplus_{p'} (x_j, \vec x; \epsilon)), \vec b, c, d) \\
 = & f(\vec x, \min (a, \oplus_{p} (\vec x; \epsilon)), \vec b, c, d)  \tpkt
\end{align*}
We show the claim by (side) induction on $|z|$.
In the base case 
$$f^1_{p'} (\epsilon, \vec x; a, \vec b, c, d) = c =
 f(\vec x, \min (a, \oplus_{p'} (z, \vec x; \epsilon)),
   \vec b, c, d)$$
since $(a, \oplus_{p'} (z, \vec x; \epsilon)) = \epsilon$.
In the induction case arguments split into three subcases.

\textsc{Subcase 1.}
$|a| \leqslant p' (|z|, |\vec x|)$:
In this case $\ominus_{p'} (z, \vec x; a) = \epsilon$ holds.
Hence the following equality holds:
\begin{align*}
f^1_{p'} (zi, \vec x; a, \vec b, c, d)
&=
f[q'] (z, \vec x; \ominus_{p'} (z, \vec x; a), \vec b,
       f^1_{p'} (z, \vec x; a, \vec b, c, d),
       \oplus_{p'} (z, \vec x; a)
      ) \\
&=
f (z, \vec x, \epsilon, \vec b,
       f^1_{p'} (z, \vec x; a, \vec b, c, d),
       \oplus_{p'} (z, \vec x; a)
      ) 
& \text{(by IH on $q$)} \\
&=
f^1_{p'} (z, \vec x; a, \vec b, c, d) \\
&=
f(\vec x, \min (a, \oplus_{p'} (z, \vec x; a)), \vec b, c, d)
& \text{(by SIH)} \\
&=
f(\vec x, \min (a, \oplus_{p'} (zi, \vec x; a)), \vec b, c, d) \tpkt
\end{align*}

\textsc{Subcase 2.}
$p' (|z|, |\vec x|) < |a| \leqslant p' (|z| +1, |\vec x|)$:
In this case 
$|\ominus_{p'} (z, \vec x; a)| = |a| - p' (|z|, |\vec x|)$ and
$\min (\ominus_{p'} (z, \vec x; a), \oplus_{p'} (z, \vec x; a)) =
 \ominus_{p'} (z, \vec x; a)$
hold.
Hence the following equality holds:
\begin{align*}
\mparbox{3mm}{f^1_{p'} (zi, \vec x; a, \vec b, c, d)} \\ 
&=
f[q'] (z, \vec x; \ominus_{p'} (z, \vec x; a), \vec b,
       f^1_{p'} (z, \vec x; a, \vec b, c, d),
       \oplus_{p'} (z, \vec x; a)
      ) \\
&=
f[q] (\vec x; \ominus_{p'} (z, \vec x; a), \vec b,
       f^1_{p'} (z, \vec x; a, \vec b, c, d),
       \oplus_{p'} (z, \vec x; a)
      ) \\
&=
f    (\vec x, 
      \min (\ominus_{p'} (z, \vec x; a), \oplus_{p'} (z, \vec x; a)), \vec b,
       f^1_{p'} (z, \vec x; a, \vec b, c, d),
       \oplus_{p'} (z, \vec x; a)
      ) 
& \text{(by IH on $q$)} \\
&=
f    (\vec x, 
      \ominus_{p'} (z, \vec x; a), \vec b,
       f^1_{p'} (z, \vec x; a, \vec b, c, d),
       \oplus_{p'} (z, \vec x; a)
      ) \\
&=
f    (\vec x, 
      \ominus_{p'} (z, \vec x; a), \vec b,
        f(\vec x; \min (a, \oplus_{p'} (z, \vec x; \epsilon)),
          \vec b, c, d),
       \oplus_{p'} (z, \vec x; a)
      ) 
& \text{(by SIH)} \\
&=
f    (\vec x, 
      \ominus_{p'} (z, \vec x; a), \vec b,
        f(\vec x; \oplus_{p'} (z, \vec x; \epsilon), \vec b, c, d),
       \oplus_{p'} (z, \vec x; a)
      ) \\
&=
f(\vec x, a, \vec b, c, d) 
& \text{(by Lemma \ref{lem:oplus})} \\
&=
f(\vec x, \min (a, \oplus_{p'} (zi, \vec x; \epsilon)),
  \vec b, c, d) \tpkt
\end{align*}

\textsc{Subcase 3.}
$p' (|z| +1, |\vec x|) < |a|$:
As in the previous subcase, the following equality holds:
\begin{align*}
\mparbox{3mm}{f^1_{p'} (zi, \vec x; a, \vec b, c, d)} \\
&=
f[q'] (z, \vec x; \ominus_{p'} (z, \vec x; a), \vec b,
       f^1_{p'} (z, \vec x; a, \vec b, c, d),
       \oplus_{p'} (z, \vec x; a)
      ) \\
&=
f[q] (\vec x; \ominus_{p'} (z, \vec x; a), \vec b,
       f^1_{p'} (z, \vec x; a, \vec b, c, d),
       \oplus_{p'} (z, \vec x; a)
      ) \\
&=
f    (\vec x, 
      \min (\ominus_{p'} (z, \vec x; a), \oplus_{p'} (z, \vec x; a)), \vec b,
       f^1_{p'} (z, \vec x; a, \vec b, c, d),
       \oplus_{p'} (z, \vec x; a)
      ) 
& \text{(by IH on $q$)} \\
&=
f    (\vec x, 
      \oplus_{p'} (z, \vec x; a), \vec b,
       f^1_{p'} (z, \vec x; a, \vec b, c, d),
       \oplus_{p'} (z, \vec x; a)
      ) \\
&=
f    (\vec x, 
      \oplus_{p'} (z, \vec x; a), \vec b,
        f(\vec x, \min (a, \oplus_{p'} (z, \vec x; \epsilon)),
          \vec b, c, d),
       \oplus_{p'} (z, \vec x; a)
      ) 
& \text{(by SIH)} \\
&=
f    (\vec x, 
      \ominus_{p'} (z, \vec x; a), \vec b,
        f(\vec x; \oplus_{p'} (z, \vec x; \epsilon), \vec b, c, d),
       \oplus_{p'} (z, \vec x; a)w
      ) \\
&=
f(\vec x, \ominus_{p'} (zi, \vec x; a), \vec b, c, d) 
& \text{(by Lemma \ref{lem:oplus})} \\
&=
f(\vec x, \min (a, \oplus_{p'} (zi, \vec x; \epsilon)),
  \vec b, c, d).
\end{align*}
This completes the case and hence the proof of the lemma.
\end{proof}

\begin{lemma}[Recursion simulation lemma]
\label{lem:recursion_simulation}
(Cf. \cite[Theorem 3.6]{HW99})
Let $f$ be an $l$-ary polynomial-time function.
Then for any polynomial 
$p: \mathbb N^k \rightarrow \mathbb N$ with
 non-negative coefficients there exists a function 
$\{ f, p\} \in \Bwsc^{k, l}$ such that for all $\vec b$, 
$\{ f, p \} (\vec x; \vec b) = f(\vec b)$ holds
whenever $\max |\vec b| \leqslant p (|\vec x|)$.
\end{lemma}

\begin{proof}
We employ the recursion-theoretic characterisation of the polynomial
 time functions by A. Cobham \cite{Cob64}.
Namely the class of all the polynomial-time computable functions
 coincides with the class $\mathcal C$, which is the
 smallest class containing the constant function 
$(x_1, \dots, x_k) \mapsto \varepsilon$, 
the successor functions 
$x \mapsto x0$ and $x \mapsto x1 $, and the projection functions 
$(x_1, \dots, x_k) \mapsto x_j$, and closed under composition and
 polynomially length-bounded recursion on notation. 
Let $f$ be a polynomial-time computable function.
We show Recursion simulation lemma by induction over the construction of
 $f$ in the Cobham class $\mathcal{C}$.
If $f$ is one of the initial functions, then the choice of the
 witnessing function $\{ f, p \}$ is clear.

\textsc{Case.}
$f$ is defined by composition from some polynomial time functions
$h, g_1, \dots, g_{l'}$ by
$f(\vec x) = h(g_1 (\vec x), \dots, g_{l'} (\vec x))$:
Let an arbitrary polynomial $p$ with non-negative coefficients be given.
Then by IH for $g_1, \dots, g_{l'}$ for each 
$j = 1, \dots, l'$ there exists a witnessing function
$\{ g_j, p\} \in \Bwsc$.
Let $p_1, \dots, p_{l'}$ be polynomials with non-negative coefficients
 such that 
$|g_j (\vec b)| \leqslant p_j (|\vec b|)$ 
for each $j = 1, \dots, l'$.
It is well known that there exists such a length-bounding polynomial for any
 polynomial time function.
Define another polynomial $q$ by
$q(\vec x) = \sum_{j=1}^{l'} p_j (p(\vec x), \dots, p(\vec x))$.
Clearly the polynomial $q$ has only non-negative coefficients.
By IH for $h$ there exists a witnessing function 
$\{ h, q \}$.
We define a function $\{ f, p\} \in \Bwsc$ by
$\{ f, p \} (\vec x; \vec b) =
 \{ h, p \} (\vec x; \{ g_1, p \} (\vec x; \vec b), \dots,
                     \{ g_{l'}, p \} (\vec x; \vec b))$.
Suppose that $\max |\vec b| \leqslant p(|\vec x|)$ holds.
Then for each $j = 1, \dots, l'$ we have
\begin{align*}
|\{ g_j, p \} (\vec x; \vec b)| & = 
|g_j (\vec b)| 
& \text{(by IH for $g_j$)} \\ 
&\leqslant
p_j (|\vec b|) \\
&\leqslant
p_j (p(|\vec x|), \dots, p(|\vec x|))
&\text{(by monotonicity of $p_j$)} \\
&\leqslant q(|\vec x|)\tpkt
\end{align*}
Hence the following equality holds:
\begin{align*}
\{ f, p \} (\vec x; \vec b) &=
\{ h, p \} (\vec x; \{ g_1, p \} (\vec x; \vec b), \dots,
                    \{ g_{l'}, p \} (\vec x; \vec b)), \\
&=
h(\{ g_1, p \} (\vec x; \vec b), \dots, 
  \{ g_{l'}, p \} (\vec x; \vec b))
& \text{(by IH for $h$)} \\
&=
h(g_1 (\vec b), \dots, g_{l'} (\vec b))
& \text{(by IH for $g_1, \dots, g_{l'}$)} \\
&=
f(\vec b)\tpkt
\end{align*}

\textsc{Case.}
$f$ is defined by polynomially length-bounded recursion on notation from some
 polynomial time functions
$g, h_0, h_1$ and some length-bounding polynomial $p_f$ by
\begin{align*} 
f(\epsilon, \vec b) &= g(\vec b), \\
f(ai, \vec b) &= h_i (a, \vec b, f(a, \vec b))
& (i = 0, 1) \\
|f(a, \vec b)| &\leqslant p_f (|a|, |\vec b|):
\end{align*}
Let an arbitrary polynomial $p$ with non-negative coefficients be given.
By IH for $g$ there exists a witnessing function 
$\{ g, p \} \in \Bwsc$.
Define a polynomial $q$ by
$q(\vec x) = p (\vec x) + p_f (p(\vec x), \dots, p(\vec x))$.
By IH for $h_0, h_1$ there exist witnessing functions 
$\{ h_i , q \} \in \Bwsc$ for each $i = 0, 1$.
In order to define a witnessing function $\{ f, p \} \in \Bwsc$ we
define auxiliary functions 
$\numseq{h_0,q}$ and $\numseq{h_1,q}$ by
$\numseq{h_i,q} (\vec x, a, \vec b, c) =
 \{ h_i, q \} (\vec x; a, \vec b, c)$
$(i = 0, 1)$.
Further we define another auxiliary function 
$\numseq{f, p}$ by
\begin{align*}
\numseq{f,p} (\vec x, \varepsilon, \vec b, c, d) &= c, \\
\numseq{f,p} (\vec x, ai, \vec b, c, d) &=
\numseq{h_i, p} (\vec x, d \oplus a, \vec b,
                 \numseq{f,p} (\vec x, a, \vec b, c, d)).
& (i = 0, 1)
\end{align*}
By definition $\numseq{f,p}$ and $\numseq{h,q}$ meet the conditions in
 the premise of Lemma \ref{lem:oplus}.
Hence by Lemma \ref{lem:f[p]} we have a function 
$\numseq{f,p} [p] \in \Bwsc$ which witnesses Lemma \ref{lem:f[p]} on 
$\numseq{f,p}$.
Now we define the function $\{ f, p \} \in \Bwsc$ by
$\{ f, p \} (\vec x; a, \vec b) =
 \numseq{f,p} [p] (\vec x; a, \vec b, 
                   \{ g, p \} (\vec x; \vec b), \epsilon)$.
Suppose that $\max (|a|, |\vec b|) \leqslant p(|\vec x|)$ holds.
We show that $\{ f, p \} (\vec x; a, \vec b) = f(\vec b)$ holds by
(side) induction on $|a|$.
In the base case the following equality holds:
\begin{align*}
\{ f, p \} (\vec x; \epsilon, \vec b) &=
\numseq{f,p}[p] (\vec x; \epsilon, \vec b, 
                 \{ g, p \} (\vec x; \vec b), \epsilon) \\
&=
\numseq{f,p} (\vec x, \min (\epsilon, \oplus_p (\vec x; \epsilon)),
              \vec b, \{ g, p \} (\vec x; \vec b), \epsilon
             )
& \text{(by Lemma \ref{lem:f[p]})} \\
&=
\numseq{f,p} (\vec x, \epsilon,
              \vec b, \{ g, p \} (\vec x; \vec b), \epsilon
             ) \\
&=
\{ g, p \} (\vec x; \vec b) \\
& = g(\vec b)
& \text{(by IH for $g$)} \\
&= f(0, \vec b) \tpkt
\end{align*}
In the induction case
$\max (|a|, |\vec b|) \leqslant p(|\vec x|)$ holds since
we are assuming that 
$\max (|a|+1, |\vec b|) \leqslant p(|\vec x|)$ holds.
Hence the following equality holds:
\begin{align*}
\mparbox{3mm}{\{ f, p \} (\vec x; ai, b)} \\
&=
\numseq{f,p}[p] (\vec x; a_i, \vec b, 
                 \{ g, p \} (\vec x; \vec b), \epsilon), \\
&=
\numseq{f,p} (\vec x, \min (|a_i|, \oplus_p (\vec x; \epsilon)),
              \vec b, \{ g, p\} (\vec x; \vec b), \epsilon)
& \text{(by Lemma \ref{lem:f[p]})} \\
&=
\numseq{f,p} (\vec x, a_i, \vec b, \{ g, p \}(\vec x; \vec b), \epsilon)
& \text{(as $\max (|a|+1, |\vec b|) \leqslant p(|\vec x|)$)} \\
&=
\numseq{h_i,q} (\vec x, a, \vec b, 
              \numseq{f,p} (\vec x, a, \vec b, 
                            \{ g, p \} (\vec x; \vec y), \epsilon)
             ) \\    
&=
\numseq{h_i,q}(\vec x, a, \vec b, 
             \numseq{f,p} (\vec x, 
                           \min (a, \oplus_p (\vec x; \epsilon)),
                           \vec b, 
                           \{ g, p \} (\vec x; \vec b), \epsilon 
                          )
            ) \\
&=
\numseq{h_i,q}(\vec x, a, \vec b, 
             \numseq{f,p}[p](\vec x; a, \vec b,
                             \{ g, p \} (\vec x; \vec b), \epsilon)
            )
& \text{(by Lemma \ref{lem:f[p]})} \\
&=
\numseq{h_i,q}(\vec x, a, \vec b, \{ f, p \} (\vec x; a, \vec b)) =
\numseq{h_i,q} (\vec x, a, \vec b, f(a, \vec b))
& \text{(by SIH)} \\
&=
\{ h_i,q \} (\vec x; a, \vec b, f(a, \vec b)) =
h_i(a, \vec b, f(a, \vec b)) 
& \text{(by IH for $h_i$)}  \\
&=
f(ai, \vec b) \tpkt
\end{align*}
This completes the case and hence the proof of the lemma.
\end{proof}

\begin{IEEEproof}[Proof of Lemma \ref{lem:Bwsc}]
Let $f$ be a $k$-ary polynomial-time computable function.
Define a $k$-ary polynomial $p$ with non-negative coefficients by
$p(\vec x) = x_1 + \cdots + x_k$.
Then there exists a function
$\{ f, p \} \in \Bwsc$ which witnesses Lemma \ref{lem:recursion_simulation}.
By the definition of the polynomial $p$,
$\max |\vec x| \leqslant p(|\vec x|)$ holds.
Hence by Lemma \ref{lem:recursion_simulation} the equality
$\{ f, p \} (\vec x; \vec x) = f(\vec x)$ holds.
This implies that 
$f \in \bigcup_{k \in \mathbb N} \Bwsc^{k,0}$.
\end{IEEEproof}

\section{A Non-Trivial Closure Property of the Polytime Functions}\label{s:spopstarps}

In this section we introduce \emph{small polynomial path order with parameter substitution} (\POPSTARSP \ for short), 
that extends clause \cref{gspop}{ia} to account for parameter substitution.

\begin{definition}\label{d:gspopps}
  Let $s$ and $t$ be terms such that $s = f(\pseq[k][l]{s})$.
  Then $s \gspopps t$ if one of the following alternatives holds.
  \begin{enumerate}
    \item\label{d:gspopps:st} $s_i \geqspopps t$ for some argument $s_i$ of $s$.
    \item\label{d:gspopps:ia} $f$ is a defined symbol, $t = g(\pseq[m][n]{t})$ 
      such that $g$ is below $f$ in the precedence and
      the following conditions hold:
      \begin{enumerate}
      \item\label{d:gspopps:ia:1} $s \nsubtermstrict t_j$ for all normal arguments $t_j$ of $t$;
      \item\label{d:gspopps:ia:2} $s \gspopps t_j$ for all for all normal arguments $t_j$ of $t$;
      \item\label{d:gspopps:ia:3} except for one argument $t_{j_0}$, 
        all arguments $t_{j}$ ($j\not=j_0$) 
        contain only function symbols below $f$ in the precedence.
      \end{enumerate}
    \item\label{d:gspopps:ts} $f$ is recursive and $t = g(\pseq[k][m]{t})$
      such that $g$ is equivalent to $f$ in the precedence and the following conditions hold:
      \begin{enumerate}
      \item $\tuple{s_1,\dots,s_k} \gspopps \tuple{t_{\pi(k)},\dots,t_{\pi(k)}}$ 
        for some permutation $\pi$ on normal argument positions;
      \item $s \gspopps t_{j}$ for all safe arguments $t_j$;
      \item all safe arguments $t_j$ contain only function symbols below $f$ in the precedence.
      \end{enumerate}
    \end{enumerate}
  Here $s \geqspopps t$ denotes that either $s$ and $t$ are equivalent or 
  $s \gspopps t$. 
  In the last clause, we use $\gspopps$ also for the product 
  extension of $\gspopps$ (modulo permutation).
\end{definition}

As evident from our experiments, parameter substitution extends the analytical power of 
\POPSTARS\ significantly. In particular, \POPSTARS\ can handle tail recursion as in 
the TRS $\RSrev$ whose rules are depicted in Fig.~\ref{fig:rev}. 
Whereas $\RSrev$ cannot be handled by \POPSTARS, it is compatible 
with $\gspopps$ as induced by the precedence
$\mrev \sp \mrevt \sp {\cons}$ where only $\mrevt$ is recursive. 

\begin{figure}[b]
  \centering
  \begin{alignat*}{4}
  \mrev(xs) & \to \mrevt(xs,\mnil) &
  \mrevt(\mnil, ys) & \to ys \\
  && \mrevt(x \cons xs, ys) & \to \mrevt(xs, x \cons ys)
  \end{alignat*}  
  \caption{Rewrite system $\RSrev$.}
  \label{fig:rev}
\end{figure}

Still \POPSTARSP~induces polynomially bounded runtime complexity in the sense of Theorem~\ref{t:spopstar}.
We emphasise that the proof requires only minor modification.
First, we verify that 
the set $\Tn$ is closed under rewriting in the sense of Lemma~\ref{l:spopstar:tnderiv}.

\begin{lemma}\label{l:spopstar:tnderiv:ps}
  Let $\RS$ be a completely defined TRS compatible with $\gspopps$.
  If $s \in \Tn$ and $s \irew t$ then $t \in \Tn$.
\end{lemma}
\begin{IEEEproof}
  The Lemma follows by a straight forward inductive argument on Definition~\ref{d:gspopps}.
\end{IEEEproof}

Further, innermost rewrite step embed into $\gspopv[k]$ in accordance to Lemma~\ref{l:embed}.

\begin{lemma}\label{l:embed:ps}
  Let $\RS$ be a completely defined TRS compatible with $\gspopps$. Let 
  $\ell \geqslant \max\{ \size{r} \mid {l \to r} \in \RS\}$.
  If $s \in \Tn$ and $s \irew t$ then $\ints(s) \gspopv[\ell] \ints(t)$.
\end{lemma}
\begin{IEEEproof}
  The proof follows the proof steps of Lemma~\ref{l:embed}. 
  The only deviation is that the application of the auxiliary
  Lemma~\ref{l:embed:aux} is replaced by the stronger statement:
  If $s \gspopps t$ then (i) $\fn(s_1\sigma, \dots, s_l\sigma) \gspopv[\size{t}] u$ for all $u \in \ints(t\sigma)$
  and further, (ii) at most one $u \in \ints(t\sigma)$ contains symbols not below $f$ in the precedence.
  Here $s = f(\pseq[l][m]{s})$ is a basic term, and $\sigma$ is a substitution that maps variables to values.
  Property (ii) follows again by straight forward inductive reasoning, 
  for Property (ii) the only new case is when
  $s \gspopps t$ follows by \cref{gspopps}{ia}.
  Consider $t = g(\pseq[k][n]{t})$ where
  $$
  \ints(t\sigma) = \gn(t_1\sigma, \dots, t_k\sigma) \append \ints(t_{k+1}\sigma) \append \cdots \append \ints(t_{k+n}\sigma)\tpkt
  $$
  Exactly as in Lemma~\ref{l:embed:aux} we verify $(\ddag)$.
  Unlike for the case \cref{gspop}{ia}, we cannot reason that the safe arguments $t_j$ of $t$ ($j = k+1,\dots,k+n$)
  are values. Instead, we use the induction hypothesis on $t_j$ to conclude 
  $\fn(l_1\sigma, \dots, l_m\sigma) \gspopv[\ell] u$ for all $u \in \ints(t_j\sigma)$.
\end{IEEEproof}

\begin{theorem}\label{t:spopstarps}
  Let $\RS$ denote a predicative recursive TRS of degree $d$. 
  Then the innermost derivation height of any basic term 
  $f(\svec{u}{v})$ is bounded by a polynomial of degree $d$ in the 
  sum of the depths of normal arguments $\vec{u}$.
\end{theorem}
\begin{IEEEproof}
  The proof goes in accordance to the proof of Theorem~\ref{t:spopstar},
  where we replace the application of Lemma~\ref{l:embed}
  and Lemma~\ref{l:spopstar:tnderiv} by the corresponding
  Lemma~\ref{l:embed:ps}
  and Lemma~\ref{l:spopstar:tnderiv:ps} respectively.
\end{IEEEproof}

The order \POPSTARSP\ is complete for the class of
polytime computable functions.
However, in order to state a stronger completeness result, we introduce
an extension of the class $\Bwsc$ according to the definition of \POPSTARSP.
Let $\Bwscps$ denote the smallest class containing $\Bwsc$ and closed
under weak safe composition ($\m{WSC}$) and \emph{safe recursion
on notation with parameter substitution} ($\m{SRN_{PS}}$) which is presented in
Fig.~\ref{fig:Bwscps}.
Then \POPSTARSP\ is complete for $\Bwscps$ in
the same sense as Theorem \ref{thm:strong_completeness}.
Here we adapt the notion of predicative recursive TRS of degree $d$
to \POPSTARSP\ in the obvious way.

\begin{figure*}
\begin{tabular}{l@{\hspace{3mm}}}
\textbf{Parameter Substitution} ($\m{SRN_{PS}}$) \\[3mm]
\qquad$f(0, \vec x; \vec y) =  g(\vec x; \vec y)$
\\
\qquad$f(S_i (; z), \vec x; \vec y) = 
  h_i (z, \vec x; \vec y, f(z, \vec x; \vec{p} (\vec z, \vec x; \vec y)))$
  ($i = 0, 1$)
\end{tabular}
\caption{Safe recursion on notation with parameter substitution}
\label{fig:Bwscps}
\end{figure*}

\begin{theorem}
For any $\Bwscps$-function $f$
there exists a confluent TRS
$\RS_f$ that that is predicative recursive of degree $d$, where $d$
equals the maximal number of nested application of ($\m{SRN_{PS}}$) in the
definition of $f$.
\end{theorem}
\begin{IEEEproof}
The proof goes in accordance to the proof of Theorem
 \ref{thm:strong_completeness}.
One will extend the TRS $\RS_{\Bwsc}$ for $\Bwsc$ to a TRS 
$\RS_{\Bwscps}$ for $\Bwscps$ by adding rules corresponding to the schema
of ($\m{SRN_{PS}}$). Clearly this schema is a syntactic extension of the
schema of ($\m{SRN}$). Hence we can replace the case of ($\m{SRN}$) by 
($\m{SRN_{PS}}$) and application of Definition \ref{d:gspop}.\ref{d:gspop:ts} by application
 of  Definition \ref{d:gspopps}.\ref{d:gspopps:ts}.
\end{IEEEproof}

\begin{corollary}
The class $\Bwsc$ is closed under predicative recursion with
parameter substitution.
\end{corollary}
\begin{IEEEproof}
By the theorem the extension of $\Bwsc$ with the schema ($\m{SRN_{PS}}$) yields
only functions that are representable by predicative recursive TRS of degree.
Thus these functions are polytime computable and due to Lemma~\ref{lem:Bwsc} contained
in $\Bwsc$.
\end{IEEEproof}


\section{Experimental Results}\label{s:exps}

The complexity analyser \TCT\ 
features a fully automatic implementation of \POPSTARS\ and \POPSTARSP.
To facilitate an efficient synthesis of a concrete order, 
we make use of the state-of-the-art SAT-solver \textsf{MiniSAT} \cite{ES03}.
The experiments were conducted on a laptop with 4Gb of RAM and 
Intel${}^\text{\textregistered}$ Core${}^\text{\texttrademark}$ i7-2620M CPU (2.7GHz).

In Table~\ref{tbl:exp1} we contrast the different orders on our testbed.
The testbed is a subset of 757 examples from the termination problem database, version 8.0%
\footnote{Available at \url{http://termcomp.uibk.ac.at/status/downloads/tpdb-8.0.tar.gz}}. 
This subset was obtained by restricting the runtime complexity problem 
set to constructor TRSs, additionally removing TRSs that are not wellformed.%
\footnote{Cf. \url{http://cl-informatik.uibk.ac.at/software/tct/experiments/lics2011} for full experimental evidence.}.

The rows of Table~\ref{tbl:exp1} reflect the assessed bounds on the innermost runtime complexity.
Additionally we annotate for each method the number of systems
that were proven in total (row {\small{\textsf{yes}}}), and the 
number of systems were a proof was not obtained (row {\small{\textsf{maybe}}}).
Since recursive path orders (with multiset status) (\emph{\MPO} for short)
encompass \LMPO~as well as (small) polynomial path orders, we also
included \MPO~for comparison.
This reveals that predicative recursion limits the power of our
techniques by roughly one fourth on our testbed.
Comparing \POPSTAR\ with \POPSTARS\ we see an increase in precision
accompanied with only minor decrease in power. Of the four systems that 
can be handled by \POPSTAR\ but not by \POPSTARS, two fail to be oriented
because \POPSTARS\ weakens the multiset status to product status, 
and two fail to be oriented because of the weakening of the composition scheme.
Compared to \LMPO, polynomial path orders loose in power as they cannot deal
with multiple recursive calls. 
Note that not all systems proven 
by \LMPO~admit polynomial (innermost) runtime complexity.
The last two columns of Table~\ref{tbl:exp1} demonstrate that parameter substitution 
almost closes the gap in power to \LMPO. 
Whether this extension is also possible for \LMPO~remains currently unknown.

\newcommand{\tm}[1]{\bf{\tiny{$\backslash$#1}}}
\renewcommand{\c}[2]{{\small{#2}}}
\newcommand{\spc}{@{\hspace{1.8mm}}}
\begin{table}[t]
  \centering
  \begin{tabular}{l @{\hspace{5pt}}r@{}l@{\hspace{3pt}}r@{}l@{\hspace{3pt}}r@{}l@{\hspace{3pt}}r@{}l@{\hspace{3pt}}r@{}l@{\hspace{3pt}}r@{}l}
    \hline
    \TOP
    bound
    & \multicolumn{2}{r}{\MPO} 
    & \multicolumn{2}{r}{\LMPO}
    & \multicolumn{2}{r}{\POPSTAR}
    & \multicolumn{2}{r}{\POPSTARS}
    & \multicolumn{2}{r}{\POPSTARP}
    & \multicolumn{2}{r}{\POPSTARSP} 
    \BOT
    \\
    \hline
    \TOP
    $\bigO(1)$ 
    &  & 
    &  & 
    &  & 
    & \c{9}{9} & \tm{0.06} 
    &  & 
    & \c{9}{9} & \tm{0.06} 
    \\
    $\bigO(n^1)$ 
    &  & 
    &  & 
    &  & 
    & \c{23}{32} & \tm{0.07} 
    &  & 
    & \c{37}{46} & \tm{0.09} 
    \\
    $\bigO(n^2)$ 
    &  & 
    &  & 
    &  & 
    & \c{6}{38} & \tm{0.09} 
    &  & 
    & \c{7}{53} & \tm{0.10} 
    \\
    $\bigO(n^3)$ 
    &  & 
    &  & 
    &  & 
    & \c{1}{39} & \tm{0.20} 
    &  & 
    & \c{1}{54} & \tm{0.22} 
    \\
    $\bigO(n^k)$ 
    &  & 
    &  & 
    & \c{43}{43} & \tm{0.05} 
    & \c{1}{39} & \tm{0.20} 
    & \c{56}{56} & \tm{0.05} 
    & \c{1}{54} & \tm{0.22} 
    \\
    \TOP 
    \textsf{yes}
    & \c{76}{76} & \tm{0.09} 
    & \c{57}{57} & \tm{0.05} 
    & \c{43}{43} & \tm{0.05} 
    & \c{39}{39} & \tm{0.07} 
    & \c{56}{56} & \tm{0.05} 
    & \c{54}{54} & \tm{0.08} 
    \\
    \BOT \textsf{maybe}
    & \c{681}{681} & \tm{0.16} 
    & \c{700}{700} & \tm{0.11} 
    & \c{714}{714} & \tm{0.11} 
    & \c{718}{718} & \tm{0.11} 
    & \c{701}{701} & \tm{0.11} 
    & \c{703}{703} & \tm{0.11} 
    \\
    \hline
  \end{tabular}
\caption{Experimental Results}
\label{tbl:exp1}
\end{table}


\section{Conclusion}

We propose a new order, the small polynomial path order \POPSTARS. 
Based on \POPSTARS, we delineate a class of rewrite systems, dubbed
systems of predicative recursion of degree $d$, such that for
rewrite systems in this class we obtain that the runtime complexity
lies in $O(n^d)$. This is a tight characterisation in the sense that
we can provide a family of systems of predicative recursion of depth $d$, such
that their runtime complexity is bounded from below by $\Omega(n^d)$.

In future work we want to integrate \POPSTARS\
with the weak dependency pair framework that lifts the dependency pair
method to the runtime complexity analysis. 
Furthermore, we aim to clarify the question whether the class of
predicative recursive TRS of degree $d$ \emph{exactly} characterise those
functions definable with $d$ nested applications of safe recursion. 
We conjecture that the answer is yes, but further research in this
direction is required.
In~\cite{Marion11} a type system for a simple imperative programming language
is proposed that induces polytime computability. The definition of this
type system is closely connected to predicative recursion.
We want to investigate whether a similar type system can be crafted on the
basis of the class $\Bwsc$ studied in this paper. Perhaps such a study
allows to certify more precise time bounds in the spirit of the class
of predicative recursive TRSs of degree $d$.

\bibliographystyle{plainnat}

\end{document}